%% file: retro_supermap.tex
\documentclass[pra,aps,floatfix,amsmath,superscriptaddress,twocolumn,longbibliography,nofootinbib,10pt]{revtex4-2}

\input{header}

\newcommand{\CPTP}{\mathrm{CPTP}}
\newcommand{\CP}{\mathrm{CP}}

\usepackage{tikz}
\usetikzlibrary{calc}

\input{tikzsetting}

\input{tikzicons}

\newcommand{\tooth}[1]{(#1.south west) |- (#1.north east) -- (#1.south east)}

\newcommand{\supermap}[3]{\tooth{#2} |- (#1.north) -| \tooth{#3} |- (#1.south east) -| (#2.south west)}

\newcommand{\hil}[1]{\spc{H}_{#1}}

\renewcommand{\red}[1]{#1}

\usepackage{fancybox,framed,array}

\begin{document}

\title{Bayesian retrodiction of quantum supermaps}
\author{Ge Bai}
\affiliation{Thrust of Artificial Intelligence, Information Hub, The Hong Kong University of Science and Technology (Guangzhou), Guangzhou 511453, China}
\affiliation{Centre for Quantum Technologies, National University of Singapore, 3 Science Drive 2, Singapore 117543}
\begin{abstract}

 The Petz map has been established as a quantum version of the Bayes' rule. It unifies the conceptual belief update rule of a quantum state observed after a forward quantum process, and the operational reverse process that recovers the final state to match the updated belief, effectively counteracting the forward process. Here, we study a higher-order generalization of the quantum Bayes' rule by considering a quantum process undergoing a quantum supermap. For a few families of initial beliefs, we show that a similar unification is possible -- the rules updating the beliefs about quantum channels can be implemented via a ``reverse'' quantum supermap, termed the retrodiction supermap. \red{The potential applications of retrodiction supermap are demonstrated with examples of improved error correction in quantum cloud computing}. Analytical solutions are provided for these families, while a recipe for arbitrary initial beliefs remains an open question.

\end{abstract}
\maketitle

\section{Introduction}

The Bayes' rule lies in the centre of logical reasoning \cite{pearl}. It tells how one updates one's belief of a random variable from indirect observations. In quantum generalizations of the Bayes' rule, the random variables correspond to quantum states, and the generalization is not straightforward due to operator non-commutativity. Various definitions of belief updates of quantum states has been proposed \cite{ozawa1997quantum,fuchs2001quantum,schack2001quantum,warmuth2005bayes,luders2006concerning,leifer2013towards,surace2022state,tsang2022generalized,parzygnat2022non,parzygnat2023time,parzygnat2023axioms}. Among those proposals, the Petz recovery map \cite{petz1,petz} is the only update rule that satisfies a set of desired properties analogous to the classical Bayes' rule \cite{parzygnat2023axioms}.

The Petz map highlights a unification of conceptual belief update and an operational reverse process \cite{BS21,AwBS}. Conceptually, it gives a rule to update the belief of the initial state of a process given observations on the final state; and operationally, it implements a retrodiction process that brings the final state back to a recovered initial state equal to the updated belief, effectively counteracting the original process.

The Petz map seems to have given a satisfactory, if not the final, answer to the quantum Bayes' rule. However, we found that its generalization from quantum states to quantum channels turned out to be non-trivial. 
Consider a quantum process that contains a few steps. For some steps, we have their exact characterization, while others are unknown, and we only have an initial belief about their behaviour. The steps may be ``hidden'' in between other steps and are not directly accessible. We aim to answer the following question: given observations of the process as a whole, how can we update our information about the unknown steps that may not be directly accessible?

\begin{figure}
     \centering
     \includegraphics[width=\linewidth]{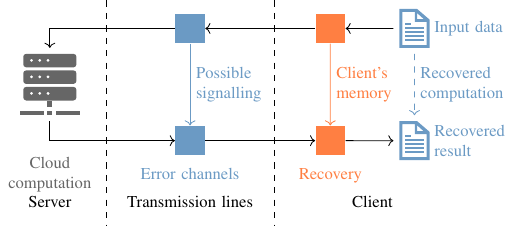}
    \caption{Application of supermap retrodiction to error correction in cloud computing. The server performs a computation procedure, and is accessed remotely via transmission lines in a quantum network. The errors in the transmission (blue boxes) are modeled as a quantum supermap acting on the computation, able to characterize possible correlations between the error channels in two directions. The client is effectively accessing a noisy version of the server's computation. To recover from the errors, the client could apply another supermap (orange boxes), consisting of correlating the original input data with the client's memory and later recovering the server's result with the help of the memory.}
    \label{fig:cloud}
\end{figure}

\red{A motivating example is quantum cloud computing \cite{arrighi2006blind,broadbent2009universal,morimae2013blind}, as illustrated in \cref{fig:cloud}. Here, the server is dedicated to applying a computation procedure to incoming data and returning the result. The client, who does not know the exact computation performed by the server, wants to apply the computation to their own data through quantum communication with the server. However, errors may occur in both directions of communication, meaning that the client effectively accesses a noisy computation with the correct computation procedure hidden in between noisy channels. This situation naturally raises the question of how one can recover the errors of the unknown process. %
This recovery can be considered as a update of client's knowledge of the server's computation via noisy communication, granting the client a more accurate computation. %
Compared with (noise-adapted) error-correction methods that considers one-way communication \cite{ng2010simple,barnum-knill,zheng2024near,biswas2024noise}, this new framework incorporates possible correlation between communication errors in both directions, and possible local quantum memory held by the client.}

\red{We formalize this problem in the framework of quantum supermaps \cite{chiribella2008transforming}.} 
A quantum supermap can be imagined as a quantum circuit board with an empty slot into which a quantum process can be embedded, as shown on the left of \cref{fig:Bayesian_network}. Such a circuit board, with all the exactly characterized steps soldered on board and leaving the unknown step as a slot, would be a supermap from the unknown step to the full quantum process.

Therefore, we call the problem of updating the belief of the unknown step ``supermap retrodiction'', in analogy to  quantum channel retrodiction that updates the belief of its input quantum state. 

\begin{figure}[t]
\begin{center}
	\begin{tikzpicture}
	\begin{scope}[xscale=0.75]
		\node[virtual] (L) at (-1.25, 0) {\phantom{$\mathcal{N}$}};
		\node[virtual] (R) at (1.25, 0) {\phantom{$\mathcal{N}$}};
		\node[virtual] (W) at (0, -0.75) {$\map{S}$};
		\draw \supermap{W}{L}{R};
		\node[tensor] (N) at (0, 0) {$\mathcal{N}$};
		\coordinate (left) at ($(L)+(-1,0)$);
		\coordinate (right) at ($(R)+(1,0)$);
		\draw (left) -- node[above]{$W$} (L) -- node[above]{$X$} (N) -- node[above]{$Y$} (R) -- node[above]{$Z$} (right);
	\end{scope}
	\begin{scope}[xshift=2.5cm,xscale=1]
		\node[draw=black, circle, minimum size=0.5cm] (W) at (0, 0) {} node at(W) {$W$};
		\node[draw=black, circle, minimum size=0.5cm, fill=black!10] (X) at (1, 0) {} node at(X) {$X$};
		\node[draw=black, circle, minimum size=0.5cm, fill=black!10] (Y) at (2, 0) {} node at(Y) {$Y$};
		\node[draw=black, circle, minimum size=0.5cm] (Z) at (3, 0) {} node at(Z) {$Z$};
		\draw[->] (W) -- (X);
		\draw[->] (X) -- (Y);
		\draw[->] (Y) -- (Z);
		\draw[->] (W) to[bend right=60] (Z);
	\end{scope}
	\end{tikzpicture}
\end{center}
	\caption{\label{fig:Bayesian_network}Quantum supermap retrodiction problem and its analogous Bayesian network. On the left, $\map{S}$ is a supermap acting on a quantum channel $\map{N}$. The supermap retrodiction aims to update one's belief on $\map{N}$, namely the correlation between systems $X$ and $Y$. On the right, it shows the Bayesian network connecting observed variables $W,Z$ and latent variables $X,Y$ with conditional probability distributions $P(X|W),P(Y|X),P(Z|YW)$. The supermap retrodiction is analogous to updating $P(Y|X)$ given observations on $W$ and $Z$.}%
\end{figure}

An analogy can be made between the supermap retrodiction and updating conditional probabilities in a Bayesian network \cite{pearl}, as shown in \cref{fig:Bayesian_network}.
In a Bayesian network, observed and latent random variables are connected by conditional probability distributions. To fit the network to observations, one needs to update the conditional probabilities from the observed variables. 
The question above, in the classical scenario, is to update the conditional probability $P(Y|X)$ involving latent variables from the observed variables $W$ and $Z$.

In this work, we propose axioms of retrodiction of supermaps, similar to those of the Petz map \cite{parzygnat2023axioms}, featuring a unification between belief update rules of quantum channels and a retrodiction supermap counteracting the original supermap. 
For general supermaps, we reduce the problem into basic cases, and give solutions for them with the reference prior chosen to satisfy some assumptions. %
However, if no assumption on the reference prior is made, even for the basic cases, finding a general solution satisfying all axioms turns out to be non-trivial. Nonetheless, we have found solutions for a few families of examples with analytical formulae to construct the retrodiction supermaps, \red{and exemplify their effectiveness for error correction in quantum cloud computing via numerical computation}.

\section{Preliminaries}

\subsection{Bayes' rule and Jeffrey's update}

Consider a stochastic map from random variable $Y$ to random variable $Z$, whose transition probabilities are denoted as $P(z|y):=P(Z=z|Y=y)$. Suppose the value of $Y$ is hidden from an observer, who wants to infer the value of $Y$ from observations of $Z$. Let $P(y|z)$ be the inferred distribution of $Y$ upon seeing $Z=z$. This can be obtained via the Bayes' rule:
\begin{align}
	P(y|z) = \frac{P(z|y)P(y)}{\sum_{y'}P(z|y')P(y')} \label{eq:clax_Bayes}
\end{align}
where $P(y):= P(Y=y)$ is a reference prior distribution, interpreted as an initial belief of the distribution of $Y$ before the observation. %

Jeffrey's rule of belief update adapts the Bayes' rule to the case of soft evidence, a blurred observation of variable $Z$ denoted as a distribution $R(z)$ \cite{jeffrey1957contributions,jeffrey1965logic,jeffrey1968probable,pearl,CHAN200567}:
\begin{align}
	Q(y) &= \sum_z  P(y|z)R(z) \\
	       &= \sum_z  \frac{P(z|y)P(y)}{P(z)} R(z)\,, \label{eq:clax_Jeffrey}
\end{align}
where $P(z)=\sum_{y'}P(z|y')P(y')$.
Jeffrey's rule signifies $P(y|z)$ as the reverse stochastic process mapping the observation $R(z)$ to the updated belief $Q(y)$. The aforementioned notion of quantum reverse process \cite{BS21,AwBS} is a generalization of this.

\subsection{Petz map}

We denote quantum systems with capital letters, and system $X$ has Hilbert space $\spc{H}_X$ and dimension $d_X$.
Let $S(\map{H})$ be the set of density operators on Hilbert space $\spc{H}$.
We denote the set of completely positive trace-preserving (CPTP) maps, namely quantum channels, from $S(\spc{H}_X)$ to $S(\spc{H}_Y)$ as $\CPTP(\spc{H}_X,\spc{H}_Y)$.

The Petz map \cite{petz1,petz} gives a general recipe for the retrodiction of a quantum process $\map{E}\in\CPTP(\spc{H}_X,\spc{H}_Y)$ and is defined as \cite{BS21,buscemi2022observational,parzygnat2023axioms}:
\begin{align} \label{eq:Petz}
	\map{R}^{\map{E},\gamma}(\sigma) := \sqrt{\gamma} \map{E}^\dag \left( \map{E}(\gamma)^{-1/2} \sigma \map{E}(\gamma)^{-1/2}  \right) \sqrt{\gamma}\,,
\end{align}
where $\gamma\in S(\spc{H}_X)$ is a reference state, corresponding to the prior belief in the Bayes' rule, and the resulting map $\map{R}^{\map{E},\gamma}$ is in $\CPTP(\spc{H}_Y,\spc{H}_X)$. We assume both $\gamma$ and $\map{E}(\gamma)$ to be full-rank for convenience.

\subsection{Quantum supermaps}

Quantum supermaps refer to transformations from one quantum process to another.
In this paper, we consider deterministic supermaps, also known as superchannels, which are completely positive linear maps transforming CPTP maps to CPTP maps \cite{chiribella2008transforming}. These objects are the higher-order counterparts of deterministic quantum processes, namely quantum channels. Given black-box access to a quantum channel $\map{N}$, one can realize $\map{S}(\map{N})$ deterministically if and only if $\map{S}$ is a superchannel \cite{chiribella2008transforming,chiribella2008quantum,chiribella2009theoretical}.

Any superchannel $\map{S} : \CPTP(\spc{H}_X,\spc{H}_Y)\to \CPTP(\spc{H}_W,\spc{H}_Z)$ acting on a CPTP map $\map{N}$ can always be realized with the structure shown in \cref{fig:supermap} \cite{chiribella2008transforming}. In equation,
\begin{align} \label{eq:supermap_decomposition}
	\map{S}(\map{N}) = \Tr_{A_R} \circ \map{U}_R \circ (\map{N}\otimes\map{I}_{A_M}) \circ \map{U}_L \circ (\map{I}_{W} \otimes \map{P}_{\ket\phi})
\end{align}
where $A_L, A_M$ and $A_R$ are ancillary systems, $\map{I}_{A_M}$ ($\map{I}_{W}$) is the identity channel on system $\spc{H}_{A_M}$ ($\spc{H}_{W}$), $\map{P}_{\ket{\phi}} \in \CPTP(\C, \spc{H}_{A_L})$ is a preparation of a fixed pure state $\ket\phi \in \spc{H}_{A_L}$, $\map{U}_L \in \CPTP(\spc{H}_W\otimes\spc{H}_{A_L},\spc{H}_X\otimes\spc{H}_{A_M})$ and $\map{U}_R\in\CPTP(\spc{H}_Y\otimes\spc{H}_{A_M},\spc{H}_Z\otimes\spc{H}_{A_R})$ are unitary channels (the dimensions satisfy $d_Wd_{A_L}=d_Xd_{A_M}, d_Yd_{A_M}=d_Zd_{A_R}$).

\begin{figure}[h]
\centering
	\begin{tikzpicture}
	\begin{scope}[xscale=0.75]
		\node[virtual] (L) at (-1.25, 0) {\phantom{$\mathcal{N}$}};
		\node[virtual] (R) at (1.25, 0) {\phantom{$\mathcal{N}$}};
		\node[virtual] (W) at (0, -0.75) {$\map{S}$};
		\draw \supermap{W}{L}{R};
		\node[tensor] (N) at (0, 0) {$\mathcal{N}$};
		\coordinate (left) at ($(L)+(-1,0)$);
		\coordinate (right) at ($(R)+(1,0)$);
		\draw (left) -- node[above]{$W$} (L) -- node[above]{$X$} (N) -- node[above]{$Y$} (R) -- node[above]{$Z$} (right);
	\end{scope}
	\node (equal) at (2,0) {$=$};
	\begin{scope}[xshift=4.5cm,xscale=0.75]
		\node[tensor] (N) at (0, 0) {$\mathcal{N}$};
		\node[virtual] (L) at (-1.25, 0) {\phantom{$\map{S}_L$}};
		\node[virtual] (R) at (1.25, 0) {\phantom{$\map{S}_R$}};
		\node[virtual] (aL) at (L|-W) {\phantom{$\map{S}_L$}};
		\node[virtual] (aR) at (R|-W) {\phantom{$\map{S}_R$}};
		\node[tensor, anchor=north, minimum height=1.25cm] (mL) at (L.north) {$\map{U}_L$};
		\node[tensor, anchor=north, minimum height=1.25cm] (mR) at (R.north) {$\map{U}_R$};
		\coordinate (left) at ($(L)+(-1.5,0)$);
		\coordinate (right) at ($(R)+(1.5,0)$);
		\draw (left) -- node[above]{$W$} (L) -- node[above]{$X$} (N) -- node[above]{$Y$} (R) -- node[above]{$Z$} (right);
		\node (x) at (aR-|right) {$\times$};
		\node (phi) at (aL-|left) {$\ket{\phi}$};
		\draw (phi) -- node[below]{$A_L$} (aL) -- node[below]{$A_M$} (aR) -- node[below]{$A_R$} (x.center);
	\end{scope}
	\end{tikzpicture}
\caption{\label{fig:supermap} The decomposition of a superchannel.}
\end{figure}

We will call $\map{U}_L$ and $\map{U}_R$ the first and the second tooth of the supermap $\map{S}$.

From \cref{fig:supermap}, one can also view $\map{S}$ as a channel from $S(\spc{H}_W\otimes\spc{H}_Y)$ to $S(\spc{H}_X\otimes\spc{H}_Z)$, and thus the Choi operator of $\map{S}$ can be defined \cite{jamiolkowski1972linear,choi1975completely,chiribella2008transforming,chiribella2008quantum,chiribella2009theoretical}.
One can use the following criterion to determine whether a map is a superchannel.

\begin{lemma}[\cite{chiribella2008transforming,chiribella2008quantum,chiribella2009theoretical}]
	$C\in S(\spc{H}_W\otimes\spc{H}_X\otimes\spc{H}_Y\otimes\spc{H}_Z)$ is the Choi operator for a superchannel $\map{S}$ if and only if
	\begin{align} \label{eq:valid_supermap}
		\left\{
		\begin{array}{rcl}
			\Tr_Z[C]  &=& \Tr_{YZ}[C] / d_Y \otimes \openone_{Y} \\
			\Tr_{XYZ}[C] &=& d_Y\openone_W
		\end{array}
		\right.
	\end{align}
\end{lemma}

The rank of a supermap $S$ is defined as the rank of its Choi operator. This rank is also the minimal dimension of $A_R$ over realizations of $S$ \cite{bisio2011minimal}.

\section{From retrodiction maps to supermaps}

For the retrodiction of superchannel $\map{S}: \CPTP(\spc{H}_X, \spc{H}_Y) \to \CPTP(\spc{H}_Z, \spc{H}_W)$, the reference prior, namely the initial belief, is a channel $\Gamma \in \CPTP(\spc{H}_X,\spc{H}_Y)$, and the retrodiction supermap aims to update the belief from the observed output of $\map{S}$. 
Let the observed output be $\map{M}\in\CPTP(\spc{H}_Z, \spc{H}_W)$. The retrodiction supermap of $\map{S}$ with reference prior $\Gamma$, denoted as $\map{R}^{\map{S},\Gamma}: \CPTP(\spc{H}_Z, \spc{H}_W) \to \CPTP(\spc{H}_X, \spc{H}_Y)$, is supposed to map $\map{M}$ to an updated belief $\map{R}^{\map{S},\Gamma}(\map{M})$.

Ref. \cite{parzygnat2023axioms} lists a set of axioms for retrodiction, which are all satisfied by the Petz map.
We take the same set of axioms, replace the channels with superchannels, and list them as follows.

\makeatletter
\newlist{axiomenum}{enumerate}{1}
\setlist[axiomenum,1]{label={\arabic*.},
                   ref = {\arabic*}}
\crefname{axiomenumi}{Property}{Properties}
\crefformat{axiomenumi}{#2\cref@axiomenumi@name~#1#3}
\crefrangeformat{axiomenumi}{#3\cref@axiomenumi@name@plural~#1#4 to #5#2#6}
\crefmultiformat{axiomenumi}{#2\cref@axiomenumi@name@plural~#1#3}{ and~#2#1#3}{, #2#1#3}{ and~#2#1#3}
\makeatother
\begin{axiomenum}
	\item \label{axiom:superchannel} Deterministic: The retrodiction supermap is a superchannel, namely it is completely positive map satisfying condition \cref{eq:valid_supermap} and thus has a deterministic implementation.
	\item \label{axiom:recover_prior} No surprise no update: If the input of the retrodiction supermap is the propagated reference prior, it should recover the reference prior. Namely,
	\begin{align} \label{eq:recover_prior}
		\map{R}^{\map{S},\Gamma}(\map{S}(\Gamma)) =\Gamma .
	\end{align}
	In other words, if the observation matches the prior belief exactly, no update will be made on the belief.

	\item \label{axiom:reversible} Recover whenever possible: If the superchannel $\map{S}$ is perfectly recoverable, namely there exists another superchannel $\map{T}$ such that $\map{T}\circ\map{S}$ is the identity supermap, then the retrodiction supermap $\map{R}^{\map{S},\Gamma}$ also satisfies that $\map{R}^{\map{S},\Gamma}\circ\map{S}$ is the identity supermap. %

	\item \label{axiom:compositional} Compositional: The retrodiction supermap for the composition of two superchannels $\map{S}_2 \circ \map{S}_1$ is the composition of their respective retrodiction supermaps in the reverse order, with priors properly propagated forward. Namely, 
	\begin{align}
		\map{R}^{\map{S}_2\circ\map{S}_1, \Gamma} = \map{R}^{\map{S}_1,\Gamma} \circ \map{R}^{\map{S}_2,\map{S}_1(\Gamma)}.
	\end{align}

	\item \label{axiom:tensorial} Tensorial: The retrodiction supermap for the tensor product of two superchannels $\map{S}_1 \otimes \map{S}_2$ is the tensor product of their respective retrodiction supermaps:
	\begin{align}
		\map{R}^{\map{S}_1\otimes \map{S}_2, \Gamma_1\otimes\Gamma_2} = \map{R}^{\map{S}_1,\Gamma_1} \otimes \map{R}^{\map{S}_2,\Gamma_2}.
	\end{align}

\end{axiomenum}

\cref{axiom:superchannel} is the key for the unification between belief update and the reverse process. Some belief update rules lack deterministic physical implementations. For states, examples include the reverse processes based on Jordan product \cite{parzygnat2023time} and Lie product formula \cite{warmuth2005bayes}, and for channels, one example is the one using Petz map and Choi--\Jami{} isomorphism to be mentioned in \cref{sec:reductions}. The inability to implement the retrodiction map deterministically  prevents their use in critical applications like error correction. Therefore, in the following, we only consider deterministically realizable retrodiction supermaps. Some bounds on the quality of recovering the prior channel with general supermaps has been studied in \cite{pandey2024fundamental}.

\cref{axiom:recover_prior,axiom:reversible} are also crucial for the retrodiction supermap, with the former highlighting the reference prior as the pivot and the latter explaining why retrodiction is considered a ``reverse'' process. %
\red{
\cref{axiom:compositional,axiom:tensorial} enable convenient simplifications to the problem, and in turn, can be used as rules to compose a retrodiction supermap from basic cases, which will be investigated in the next section.}

\red{There is one more axiom listed in Ref. \cite{parzygnat2023axioms}, whose supermap counterpart can be taken as:}
\red{
\begin{axiomenum}[resume*]
	\item \label{axiom:involutive} Involutive: If $\map{R}^{\map{S},\Gamma}$ is a retrodiction supermap for $\map{S}$ with prior $\Gamma$, then $\map{S}$ is a retrodiction supermap for $\map{R}^{\map{S},\Gamma}$ with prior $\map{S}(\Gamma)$. Namely,
	\begin{align}
		\map{R}^{\map{R}^{\map{S},\Gamma},\map{S}(\Gamma)} = \map{S}.
	\end{align}
\end{axiomenum}
}
\red{In the original context of quantum channels, this property implies that the retrodiction can be interpreted as a ``time reversal'' of the original process. However, in the supermap case, its relation to time reversal is not as clear since the operations belonging to the original and retrodiction supermaps are interleaved. 
\cref{axiom:involutive} will not be strictly imposed in the following due to technical difficulty, and the difficulty may come from the inability to correspond the retrodiction supermap   to a time reversal.}

\begin{table*}
\caption{\label{tab:S}Summary of the retrodiction for $\map{S}_1,\map{S}_2,\map{S}_3,\map{S}_4$. The empty boxes indicates the input channels of supermaps.}
\newcommand{\centered}[1]{\begin{tabular}{l} #1 \end{tabular}}
	\begin{tabular}{|c|lc|c|c|}
	\hline
		~ & \multicolumn{2}{c|}{Effect of the supermap} & \centered{Recoverable} & Retrodiction supermap\\
	\hline
		$\map{S}_1$ & \centered{Tensor product \\ with identity} & \centered{\begin{tikzpicture} \node at(0,0.25) {}; \node[tensor] (M) at (0, 0) {}; \draw (-0.75,0) -- (M) -- (0.75,0); \draw(-0.75,-0.5) -- (0.75,-0.5);\end{tikzpicture}} & Yes & Ignore the added channel\\
	\hline
		$\map{S}_2$ & \centered{Prepending and \\ appending unitaries} & \centered{\begin{tikzpicture} \node at(0,0.25) {};\node[tensor](L)at(-0.75,0){$U_L$}; \node[tensor](M)at(0,0){}; \node[tensor](R)at(0.75,0){$U_R$}; \draw(-1.25,0)--(L)--(M)--(R)--(1.25,0);\end{tikzpicture}} & Yes & \centered{\begin{tikzpicture} \node at(0,0.25) {};\node[tensor](L)at(-0.75,0){$U_L^\dag$}; \node[tensor](M)at(0,0){}; \node[tensor](R)at(0.75,0){$U_R^\dag$}; \draw(-1.25,0)--(L)--(M)--(R)--(1.25,0);\end{tikzpicture}}\\
	\hline
		$\map{S}_3$ & \centered{Fixing part \\ of the input} & \centered{\begin{tikzpicture}
			\node at(0,0.5) {};
			\node[tensor2h](M)at(0,0){};
			\coordinate (upper) at (0, 0.25);
			\node[anchor=east](s)at(-0.5,-0.25){$\ket\phi$};
			\draw (s) -- (M.west|-s);
			\draw (-0.5, 0.25) -- (M.west|-upper) (M.east|-upper) -- (0.5,0.25);
		\end{tikzpicture}}
		& No & \multicolumn{1}{l|}{\centered{Classically update only the observed subset,\\ quantumly non-trivial }} \\
	\hline
		$\map{S}_4$ & \centered{Discarding part \\ of the output} & \centered{\begin{tikzpicture}
			\node at(0,0.5) {};
			\node[tensor2h](M)at(0,0){};
			\coordinate (upper) at (0, 0.25);
			\node(s)at(0.5,-0.25){$\times$};
			\draw (s.center) -- (M.east|-s);
			\draw (-0.5, 0.25) -- (M.west|-upper) (M.east|-upper) -- (0.5,0.25);
		\end{tikzpicture}}
		& No & \multicolumn{1}{l|}{\centered{Classically use conditional Bayes' rule,\\ quantumly non-trivial }} \\
	\hline
	\end{tabular}
\end{table*}

\section{Basic cases of supermap retrodiction} \label{sec:reductions}

Due to the Choi--\Jami{} isomorphism \cite{jamiolkowski1972linear,choi1975completely}, transformations on channels can be viewed as transformations on their Choi operators. Indeed, a superchannel $\map{S} : \CPTP(\spc{H}_X,\spc{H}_Y)\to \CPTP(\spc{H}_W,\spc{H}_Z)$ defines a completely positive (CP) mapping between Choi operators \cite{chiribella2008transforming}. We denote this map as 
\begin{align}
&\map{C}_{\map{S}} \in \CP(\spc{H}_X\otimes\spc{H}_Y,\spc{H}_W\otimes\spc{H}_Z) \nonumber \\
&\map{C}_{\map{S}}:C_{\map{N}} \mapsto C_{\map{S}(\map{N})} , \label{eq:CP_rep}
\end{align}
where $\CP$ denotes the set of completely positive maps. This CP map and the Choi operator appearing in \cref{eq:valid_supermap} are different but equivalent representations of the same supermap $\map{S}$.

One may think about defining the retrodiction supermap $\map{R}^{\map{S},\Gamma}$ via the retrodiction of $\map{C}_{\map{S}}$, such that $\map{C}_{\map{R}^{\map{S},\Gamma}}$ is the Petz map of $\map{C}_{\map{S}}$ with prior $C_{\Gamma}$.
Unfortunately, the Petz map does not always give a valid superchannel satisfying \cref{axiom:superchannel}. This can be seen from the example in \cref{app:Petz_not_superchannel}. Nevertheless, we will show in later sections how the Petz map is helpful in some cases.

The aforementioned properties are the desiderata of retrodiction supermaps we aim to satisfy. They also allow us to reduce the original problem to simpler cases. 

According to the decomposition in \cref{eq:supermap_decomposition,fig:supermap}, we can write a supermap $\map{S}$ as $\map{S} = \map{S}_4 \circ\map{S}_3 \circ\map{S}_2 \circ\map{S}_1$, with
\begin{align}
	&\map{S}_1: \map{N} \mapsto \map{N}\otimes\map{I}_{A_M} \\
	&\map{S}_2: \map{N} \mapsto \map{U}_R \circ \map{N} \circ \map{U}_L \\
	&\map{S}_3: \map{N} \mapsto \map{N} \circ (\map{I}_W\otimes\map{P}_{\ket{\phi}})\\
	&\map{S}_4: \map{N} \mapsto \Tr_{A_R} \circ \map{N}
\end{align}

By the compositional property (\cref{axiom:compositional}), to obtain the retrodiction supermap for $\map{S}$, one may construct the retrodiction supermaps for $\map{S}_1,\dots,\map{S}_4$ and compose them as
\begin{align}
	&\map{R}^{\map{S},\Gamma} = \nonumber \\
	&\map{R}^{\map{S}_1,\Gamma} \circ \map{R}^{\map{S}_2,\map{S}_1(\Gamma)} \circ \map{R}^{\map{S}_3,(\map{S}_2\circ\map{S}_1)(\Gamma)} \circ \map{R}^{\map{S}_4,(\map{S}_3,\map{S}_2\circ\map{S}_1)(\Gamma)}
\end{align}

Next, we discuss separately the retrodiction supermaps for $\map{S}_1,\dots,\map{S}_4$, which are summarized in \cref{tab:S}.

\subsection{Retrodiction of $\map{S}_1$.}
	The mapping $\map{S}_1: \map{N} \mapsto \map{N}\otimes\map{I}_{A_M}$ is a recoverable superchannel. Its effect can be undone by simply ignoring the added identity channel, obtaining the retrodiction supermap. One can also justify this from the axioms.

	$\map{S}_1$ can be decomposed as the tensor product of two superchannels: $\map{S}_1 = \mathrm{id} \otimes \map{P}_{\map{I}_{A_M}}$, where $\mathrm{id}$ is the identity supermap $\map{N}\mapsto\map{N}$, and $\map{P}_{\map{I}_{A_M}}$ is a supermap that prepares $\map{I}_{A_M}$, namely it receives no input (its domain is $\CPTP(\C,\C) = \{1\}$) but outputs $\map{I}_{A_M}$.

	By \cref{axiom:reversible}, the retrodiction supermap for $\mathrm{id}$ is simply itself. The retrodiction supermap for $\map{P}_{\map{I}_{A_M}}$, denoted as $\map{R}^{\map{P}_{\map{I}_{A_M}},1}$, is a superchannel whose range is $\CPTP(\C,\C) = \{1\}$. $\map{R}^{\map{P}_{\map{I}_{A_M}},1}$ thus must ignore the input channel and produce nothing (giving the input channel an arbitrary state and discarding the output). 
	By \cref{axiom:tensorial}, $\map{R}^{\map{S}_1,\Gamma} = \mathrm{id} \otimes \map{R}^{\map{P}_{\map{I}_{A_M}},1}$, a supermap that ignores system $A_M$.

\subsection{Retrodiction of $\map{S}_2$.} Since both $\map{U}_L$ and $\map{U}_R$ are unitary, the mapping $\map{S}_2$ is one-to-one, and its inverse is $\map{S}_2^{-1}: \map{N} \mapsto \map{U}_R^\dag \circ \map{N} \circ \map{U}_L^\dag$. Thus, by \cref{axiom:reversible}, its retrodiction supermap $\map{R}^{\map{S}_2,\Gamma}$ satisfies $\map{R}^{\map{S}_2,\Gamma} \circ \map{S}_2 = \mathrm{id}$ and must be equal to $\map{S}_2^{-1}$. In this case, the retrodiction supermap is unique and does not depend on the choice of the prior $\Gamma$.

\subsection{Retrodiction of $\map{S}_3$.} \label{sec:S3_initial}
For simplicity of presentation, we relabel the systems according to \cref{fig:S3}. $\map{S}_3: \CPTP(\hil{W}\otimes\hil{A},\hil{Y})\to\CPTP(\hil{W},\hil{Y})$ is not recoverable, since for any $\ket{\phi^\perp}$ perpendicular to $\ket\phi$, applying $\map{S}_3$ on $\map{N}$ completely hides the values of $\map{N}(\rho \otimes \ketbrasame{\phi^\perp})$ for all $\rho\in S(\hil{W})$. To obtain some clue for this case, we consider a classical analogy.

	\begin{figure}[h]
	\begin{tikzpicture}
		\begin{scope}[xscale=0.75]
			\coordinate (W) at (0, -0.75);
			\node[virtual] (R) at (1.375, 0) {\phantom{$\map{N}$}};
			\node[virtual] (aR) at (R|-W) {\phantom{$\map{N}$}};
			\node[tensor, anchor=north, minimum height=1.25cm] (mR) at (R.north) {$\map{N}$};
			\coordinate (left) at ($(R)+(-1.25,0)$);
			\coordinate (right) at ($(R)+(1.25,0)$);
			\coordinate (below) at (0, -1.2);
			\coordinate (Y) at ($(R)+(0.75,0)$);
			\node[virtual, anchor=east] (phi) at (left|-aR) {$\ket\phi$};
			\draw (left) -- node[above]{$W$} (R) -- node[above]{$Y$} (right);
			\draw (phi) -- node[above]{$A$} (aR);
		\end{scope}
	\end{tikzpicture}
	\caption{\label{fig:S3}The supermap $\map{S}_3$.}
	\end{figure}

	We consider random variables $Y,W,A$ related by a conditional distribution with prior belief $P(y|wa)$. The effect of $\map{S}_3$ corresponds to observing the process from $W$ to $Y$ while fixing the the value of $A$. Let the observation be $R(y|w)$ and the fixed value be $a_0$.
	According to the principles of Jeffrey's update, the updated belief should match the observations and keeping the minimal deviation from the prior belief. This decides the updated belief $Q(y|wa)$. If $a=a_0$, $Q(y|wa) = R(y|w)$, and if $a\neq a_0$, $Q(y|wa) = P(y|wa)$. Writing $Q$ as a linear function of $R$, we obtain
	\begin{align} \label{eq:S3_clax}
		Q(y|wa) = \delta_{aa_0} R(y|w) + (1-\delta_{aa_0})P(y|wa)\,,
	\end{align}
	where $\delta_{aa_0}$ is the Kronecker delta. However, a generalization of this update to the quantum case poses constraints to the prior channel $\Gamma$ one could choose. %

In the quantum case, assuming the retrodiction supermap $\map{R}^{\map{S}_3,\Gamma}$ is applied to $\map{M}\in\CPTP(\hil{W},\hil{Y})$, one may update the belief similarly with the following procedure
	\begin{enumerate}
		\item If system $A$ is in state $\ket\phi$, apply $\map{M}$ on system $W$;
		\item If system $A$ is in a state orthogonal to $\ket\phi$, apply $\Gamma$ on the joint system $WA$.
	\end{enumerate}
	However, the classical analogy doesn't tell us what to do when $A$ contains coherence between the subspace of $\ket\phi$ and its orthogonal complement. Nevertheless, if the prior $\Gamma$ is chosen to destroy the coherence between the two subspaces, the retrodiction supermap can be defined following the classical formula \cref{eq:S3_clax}.

	The idea is to make a weak measurement on system $W$ and decide the application of $\map{M}$ or $\Gamma$ according to the outcome. The weak measurement is characterized by a quantum instrument $\{\map{J}_1,\map{J}_0\}$ defined as
	\begin{align}
		\map{J}_1(\rho) &:=\ketbrasame{\phi}\rho\ketbrasame{\phi}\\
		\map{J}_0(\rho) &:= (\openone_A-\ketbrasame{\phi})\rho(\openone_A-\ketbrasame{\phi})
	\end{align}
	$\map{J} := \map{J}_1 + \map{J}_0$ is a quantum channel that destroys coherence between the subspace of $\ket\phi$ and its orthogonal complement. Then, we can write the retrodiction supermap as
	\begin{align}
		&\map{R}^{\map{S}_3,\Gamma}(\map{M}) := \map{M} \otimes (\Tr_A\circ\map{J}_1) + \Tr[\map{M}(\tau)] \Gamma\circ(\map{I}_W\otimes\map{J}_0) \label{eq:R_S3}
	\end{align}
	where $\tau\in S(\hil{W})$ is an arbitrary state, and $\Tr[\map{M}(\tau)]$ is added to make $\map{R}^{\map{S}_3,\Gamma}$ a homogeneous linear map. An implementation of this supermap is shown in \cref{fig:S3_retro}, from which one can see that $\map{R}^{\map{S}_3,\Gamma}$ is a superchannel satisfying \cref{axiom:superchannel}.

	\begin{figure}[h]
		\begin{tikzpicture}
			\coordinate (leftouter) at (-2.75, 0);
			\coordinate (left) at (-1.75, 0);
			\coordinate (cs1) at (-1, 0);
			\node[tensor] (M) at (0, 0) {$\map{M}$};
			\coordinate (cs2) at (1, 0);
			\coordinate (right) at (1.75, 0);
			\coordinate (rightouter) at (2.75, 0);
			\coordinate (wireW) at (0, -0.75);
			\coordinate (wireA) at (0, -1.5);
			\node[tensor, minimum height=1.25cm] (Gamma) at ($(wireW)!0.5!(wireA)$) {$\Gamma$};
			\coordinate (wireC) at (0, -2.25);
			\node[prepare] (tau) at (left|-M) {$\tau$};
			\node (times) at (right|-M) {$\times$};
			\node[tensor] (J) at (left|-wireA) {$\map{J}_k$};
			\node (times2) at (right|-wireC) {$\times$};
			\draw (tau) -- (M) -- (times.center);
			\draw[double] (J) -- (J|-wireC) -- (times2.center);
			\draw (leftouter|-wireW) -- node[above]{$W$} (J.west|-wireW) -- (Gamma.west|-wireW);
			\draw (Gamma.east|-wireW) -- (right|-wireW) -- +(0.25,0) -- node[above](Ylabel){$Y$} (rightouter|-wireW);
			\draw (leftouter|-wireA) -- node[above]{$A$} (J) -- (Gamma.west|-wireA);
			\draw[fill=black] (cs1|-wireC) circle (0.05) -- (cs1|-wireW) node {$\times$} -- (cs1|-M) node {$\times$};
			\draw[fill=black] (cs2|-wireC) circle (0.05) -- (cs2|-wireW) node {$\times$} -- (cs2|-M) node {$\times$};
			\node[minimum width=1.5cm, minimum height=0.6cm] (tooth1) at ($(cs1|-M)!0.5!(tau)$){};
			\node[minimum width=1.5cm, minimum height=0.6cm] (tooth2) at ($(cs2|-M)!0.5!(times)$){};
			\node[minimum height=2.125cm] (base) at ($(wireW)!0.5!(wireC) + (0,0.0625)$) {};
			\draw[dashed] \supermap{base}{tooth1}{tooth2};
			\node[anchor=west] at (Ylabel.west|-wireC) {$\map{R}^{\map{S}_3,\Gamma}$};
		\end{tikzpicture}
		\caption{\label{fig:S3_retro} Possible implementation of the retrodiction supermap $\map{R}^{\map{S}_3,\Gamma}$ in \cref{eq:R_S3} acting on $\map{M}$. The double line denotes the classical system storing the measurement outcome of the instrument $\{\map{J}_1,\map{J}_0\}$ (denoted as $\map{J}_k$ in the figure). The controlled-SWAP gates are activated if the outcome is the one corresponding to $\map{J}_1$. The components of $\map{R}^{\map{S}_3,\Gamma}$ are in the dashed frame.}
	\end{figure}

	With the assumption that $\Gamma$ destroys the coherence, namely
	\begin{align}
		\Gamma \circ \left(\map{I}_W \otimes \map{J}\right) =\Gamma \,, \label{eq:Gamma_destroys_coherence_S3}
	\end{align}
	where $\map{J}:=\map{J}_1+\map{J}_0$, \cref{axiom:recover_prior} is also satisfied:
	\begin{align}
		& \mathrel{\phantom{=}}\map{R}^{\map{S}_3,\Gamma}(\map{S}_3(\Gamma)) \nonumber \\
		& = \Gamma \circ (\map{I}_W\otimes \map{P}_{\ket\phi} \circ\Tr_A\circ\map{J}_1) +  \Gamma\circ(\map{I}_W\otimes\map{J}_0) \\
		& = \Gamma \circ  \left(\map{I}_W\otimes(\map{J}_1 + \map{J}_0)\right)\\
		& = \Gamma \circ  \left(\map{I}_W\otimes\map{J}\right)\\
		& =\Gamma
	\end{align}
noticing that $\map{P}_{\ket\phi} \circ\Tr_A\circ\map{J}_1 = \map{J}_1$.

We have obtained the retrodiction superchannel of $\map{S}_3$ for a subset of reference priors satisfying \cref{eq:Gamma_destroys_coherence_S3}. Thus, this construction satisfies \cref{axiom:superchannel,axiom:recover_prior}. %
However, \cref{axiom:involutive} cannot be verified at this stage since the retrodiction of $\map{R}^{\map{S}_3,\Gamma}$ requires the solutions of all the basic cases ($\map{S}_1,\dots,\map{S}_4$), which we do not have full answers yet. Nonetheless, we believe this is a good candidate since it closely follows from the classical belief update rule.

\subsection{Retrodiction of $\map{S}_4$.}
For simplicity, we relabel the systems as in \cref{fig:S4} and write $\map{S}_4: \CPTP(\hil{X},\hil{Z}\otimes\hil{A})\to\CPTP(\hil{X},\hil{Z})$. We notice that this case is classically analogous to a conditional Bayes' rule. Consider one has a prior belief of a classical process $P(za|x)$, also shown in \cref{fig:S4}, corresponding to $\Gamma\in\CPTP(\spc{H}_{X},\spc{H}_{Z}\otimes\spc{H}_A)$, and one is given the observation $R(z|x)$, corresponding to $\map{S}(\map{N})$.

	\begin{figure}[h]
		\centering
	\begin{tikzpicture}
		\begin{scope}[xscale=0.75]
			\node[virtual] (R) at (1.375, 0) {\phantom{$\map{N}$}};
			\node[virtual] (aR) at (R|-W) {\phantom{$\map{N}$}};
			\node[tensor, anchor=north, minimum height=1.25cm] (mR) at (R.north) {$\map{N}$};
			\coordinate (left) at ($(R)+(-1.25,0)$);
			\coordinate (right) at ($(R)+(1.5,0)$);
			\coordinate (right2) at ($(right)+(0.5,0)$);
			\coordinate (below) at (0, -1.2);
			\coordinate (Y) at ($(R)+(0.75,0)$);
			\draw (left) -- node[above]{$X$} (R) -- node[above]{$Z$} (right) -- (right2);
			\node (x) at (aR-|right) {$\times$};
			\draw (aR) -- node[above](A){$A$} (x.center);
			\draw[dashed] ($(R.north east-|A)+(-0.3,0.2)$) rectangle ($(aR.south east-|A)+(0.3,-0.0)$);
			\node (Ylabel) at (below-|A) {$Y$};
		\end{scope}
		\begin{scope}[xshift=4cm, yshift=0.1cm, xscale=1]
			\node[draw=black, circle, minimum size=0.5cm] (X) at (0, -0.5) {} node at(X) {$X$};
			\node[draw=black, circle, minimum size=0.5cm] (Z) at (1, 0) {} node at(Z) {$Z$};
			\node[draw=black, circle, minimum size=0.5cm] (A) at (1, -1) {} node at(A) {$A$};
			\draw[->] (X) -- (Z);
			\draw[->] (X) -- (A);
			\draw[draw=black, dashed] ($(Z)!0.5!(A)$) ellipse[x radius=0.5cm, y radius=1cm];
			\node (Y) at (1.7, -0.5) {$Y$};
		\end{scope}
	\end{tikzpicture}
	\caption{\label{fig:S4}The supermap $\map{S}_4$ and the Bayesian network of its classical analogy.}
	\end{figure}

	We define the random variable $Y$ as the combination of $Z$ and $A$, that is $y=(z,a)$. To update the belief $P(y|x)=P(za|x)$, one could apply \cref{eq:clax_Bayes} with every probability distribution being conditioned on $x$, and obtain the updated belief $Q(y|x)$ as
	\begin{align}
		Q(y|x)&= \sum_{z'} P(y|z'x)R(z'|x) \\
		&= \sum_{z'}\frac{P(z'|yx)P(y|x)}{P(z'|x)} R(z'|x) \\
		&= \frac{P(y|x)}{P(z|x)} R(z|x)
	\end{align}
	This gives $Q(za|x) = R(z|x) P(za|x) / P(z|x) $, with $P(z|x)=\sum_{a'}P(za'|x)$. The last equation uses $P(z'|yx)=\delta_{zz'}$, because $z$ is determined by $y$. However, the retrodiction of $\map{S}_4$ turns out to be non-trivial in the quantum case, and will be addressed in the following and in \cref{sec:examples}.

Similar to $\map{S}_3$, the classical analogy does not directly generalized to the quantum case, but could give solutions with a restricted set of prior channels. Specifically, we consider the case where $\Gamma$ is a measure-and-prepare channel that can be written as
\begin{align}
	\Gamma(\rho) = \sum_k \Tr[\Pi_k \rho] \gamma_k ,  ~\text{with }\Pi_k\Pi_l = \delta_{kl}\Pi_k\,, \label{eq:Gamma_measure_and_prepare}
\end{align}
where the positive operator-valued measure (POVM) $\{\Pi_k\}$ is a projective measurement whose elements are projectors orthogonal to each other. Let $\{\map{J}_k\}$, defined as $\map{J}_k(\rho) := \Pi_k \rho \Pi_k$ be the quantum instrument corresponding to the projections $\{\Pi_k\}$ and
\begin{align}
	\map{J}(\rho) := \sum_k\map{J}_k(\rho) = \sum_k \Pi_k \rho \Pi_k
\end{align}
be the channel destroying the coherence between the support of different $\Pi_k$. Then, one has
\begin{align}
	\Gamma\circ\map{J} = \Gamma \,. \label{eq:Gamma_circ_J_is_Gamma}
\end{align}
We will show that in this case, one can use $W$ as a control system to selectively apply Petz maps, giving a retrodiction supermap satisfying \cref{axiom:superchannel,axiom:recover_prior}. 
This construction is shown in \cref{fig:select_petz_S3}.
\begin{figure}[h]
	\centering
	\begin{tikzpicture}
		\begin{scope}[xscale=0.75]
			\node[virtual] (R) at (1.375, 0) {\phantom{$\map{N}$}};
			\node[virtual] (aR) at (R|-W) {\phantom{$\map{N}$}};
			\node[virtual] (Rec) at (4, 0) {$\phantom{\map{R}^{\Tr_A,\gamma_k}}$};
			\node[virtual] (aRec) at (Rec|-W) {$\phantom{\map{R}^{\Tr_A,\gamma_k}}$};
			\node[tensor, anchor=north, minimum height=1.25cm] (mR) at (R.north) {$\map{N}$};
			\node[tensor, anchor=north, minimum height=1.25cm] (mRec) at (Rec.north) {$\map{R}^{\Tr_A,\gamma_k}$};
			\node[tensor] (J) at ($(R)+(-1.25,0)$) {$\map{J}_k$};
			\coordinate (left) at ($(J.west)+(-1,0)$);
			\coordinate (right) at ($(R)+(1.5,0)$);
			\coordinate (right2) at ($(Rec.east)+(0.75,0)$);
			\coordinate (below) at (0, -1.2);
			\coordinate (wireC) at (0, -1.2);
			\coordinate (Y) at ($(R)+(0.75,0)$);
			\node (x) at (aR-|right) {$\times$};
			\draw (aR) -- node[above](A){$A$} (x.center);
			\draw (left) -- node[above]{$X$} (J) -- (R) -- node[above](zlabely){\phantom{$Z$}} (Rec) -- (right2);
			\node at (A|-zlabely) {$Z$};
			\draw (aRec) -- (right2|-aRec);
			\node at ($(right2|-aRec)!0.5!(right2|-Rec)$) {$Y_{\rm r}$};
			\draw[double] (J) -- (J|-wireC) -| (aRec);
			\node[minimum size=0.6cm] (J_expand) at (J) {};
			\node[minimum height=0.3cm] (S_expand) at ($(wireC)+(1,0)$) {};
			\node[minimum height=0.6cm, minimum width=1.4cm] (Rec_expand) at (Rec) {};
			\draw[dashed] \supermap{S_expand}{J_expand}{Rec_expand};
			\node at ($(aRec)+(0,-0.8)$) {$\map{R}^{\map{S}_3,\Gamma}$};
		\end{scope}
	\end{tikzpicture}
	\caption{\label{fig:select_petz_S3} The implementation of the retrodiction supermap $\map{R}^{\map{S}_3,\Gamma}$ (dashed frame) applied on $\map{S}_3(\map{N})$. The retrodiction supermap applies a weak measurement on $X$ using the instrument $\{\map{J}_k\}$ and, at the output side, according to the measurement outcome $k$, apply $\map{R}^{\Tr_A,\gamma_k}$, the Petz map of $\Tr_A$ with reference prior $\gamma_k$ (\ref{eq:Gamma_measure_and_prepare}). $Y_{\rm r}$ denotes the recovered system $Y$.}
\end{figure}

The constructed supermap $\map{R}^{\map{S}_3,\Gamma}$ applies instrument $\{\map{J}_k\}$ on system $X$ and obtain the outcome $k$. Since $\Gamma$ is a measure-and-prepare channel satisfying \cref{eq:Gamma_circ_J_is_Gamma}, the effect of $\Gamma$ is not affected by this instrument, and the state prepared by $\Gamma$ must be $\gamma_k$. Therefore, by the property of the Petz map $\map{R}^{\Tr_A,\gamma_k}$, one has $\map{R}^{\Tr_A,\gamma_k}(\Tr_A[\gamma_k]) = \gamma_k$ and the final state on $Y_{\rm r}$ will be $\gamma_k$, consistent with $\Gamma$. Formally, given the state $\rho \in S(\hil{X})$ on system $X$, the final state on $Y_{\rm r}$ is
\begin{align}
&\mathrel{\phantom{=}}\sum_k (\map{R}^{\Tr_A,\gamma_k} \circ \Tr_A \circ \map{N} \circ \map{J}_k)(\rho) \nonumber\\
&=\sum_k (\map{R}^{\Tr_A,\gamma_k} \circ \Tr_A)\left(\sum_{k'} \gamma_{k'} \Tr[\Pi_{k'} \Pi_k \rho \Pi_k]\right)\\
&=\sum_k (\map{R}^{\Tr_A,\gamma_k} \circ \Tr_A)\left(\sum_{k'} \gamma_{k'} \Tr[\delta_{kk'}\Pi_k \rho ]\right)\\
&=\sum_k \Tr[\Pi_k \rho ] (\map{R}^{\Tr_A,\gamma_k} \circ \Tr_A)(\gamma_k)\\
&=\sum_k \Tr[\Pi_k \rho ] \gamma_k\\
& = \Gamma(\rho)
\end{align}
This shows $\map{R}^{\map{S}_3,\Gamma}(\map{S}_3(\Gamma)) = \Gamma$ and \cref{axiom:recover_prior} is satisfied. \cref{axiom:superchannel} is satisfied by construction.

\section{Examples for retrodiction of partial trace supermap with general priors} \label{sec:examples}

Although we have reduced the retrodiction problem to relatively simple supermaps in \cref{sec:reductions}, finding retrodiction supermaps for general prior $\Gamma$ satisfying the axioms turned out to be more difficult.
In this section, we focus on the partial trace supermap $\map{S}_4$. We make further analysis and provide families of cases where we have found explicit solutions of retrodiction supermaps, no longer restricted to the special case of \cref{eq:Gamma_measure_and_prepare}.
\red{In addition, we compare the effect of the supermap retrodiction with the standard Petz map in the cloud computing example for correcting the qubit-loss error induced by $\map{S}_4$. Numerical results show that the the supermap retrodiction strategy performs better in most cases even though the standard Petz map strategy is given additional information about the client's input.}

\red{In the following, we will focus on solutions satisfying \cref{axiom:superchannel,axiom:recover_prior}. This is because, \cref{axiom:reversible} is not applicable to $\map{S}_4$, and \cref{axiom:compositional,axiom:tensorial} corresponds to the composing rules of multiple supermaps, which is out of the scope of the case study here.} %

There is a trivial solution of retrodiction supermap satisfying \cref{axiom:superchannel,axiom:recover_prior}, which is a superchannel mapping every channel to $\Gamma$. %
This is not what we desire since it ignores the observations and does not update the prior belief in any case. To find non-trivial solutions, additional constraints are needed.

The additional constraints are inspired by the property of the Petz map.
From the expression of the Petz map \cref{eq:Petz}, the rank of the Petz map $\map{R}^{\map{E},\gamma}$ is never larger than that of the original map $\map{E}$\footnote{This is because their Choi operators satisfy $C_{\map{R}^{\map{E},\gamma}} = (\map{E}(\gamma)^{-1/2}\otimes\gamma^{1/2}) C_{\map{E}}^T (\map{E}(\gamma)^{-1/2}\otimes\gamma^{1/2})$ up to reordering of systems, and thus $\mathrm{rank}(C_{\map{R}^{\map{E},\gamma}}) \leq \mathrm{rank}(C_{\map{E}})$}. We make a similar constraint here, by imposing the retrodiction superchannel $\map{R}^{\map{S},\Gamma}$ to have rank no larger than $d_{A}$, which is the rank of $\map{S}_4$. When $\Gamma$ and $\map{S}_4(\Gamma)$ are full-rank, $d_{A}$ is the minimal rank of a superchannel that \cref{axiom:recover_prior} may be satisfied (shown in \cref{lem:rec_from_unitary}).
Intuitively, this minimal rank constraint requires the $\map{R}^{\map{E},\gamma}$ to keep the most information from the observation $\map{S}_4(\map{N})$. 
As we will see later in this section, this constraint also gives a form of retrodiction maps, which is similar to rotated Petz maps \cite{fawzi2015quantum,wilde2015recoverability,sutter2016universal}.

From now on, we will omit the subscript and write $\map{S} = \map{S}_4: \map{N}\mapsto \Tr_A\circ\map{N}$.

\subsection{Structure of retrodiction supermaps}
According to the decomposition in \cref{eq:supermap_decomposition,fig:supermap}, the retrodiction superchannel $\map{R}^{\map{S},\Gamma}\in\CPTP(\spc{H}_X, \spc{H}_Z) \to \CPTP(\spc{H}_X, \spc{H}_Z\otimes\spc{H}_A)$ has the structure in \cref{fig:ULUR}, where $\map{U}_L: \rho\mapsto U_L\rho U_L^\dag$, $U_L: \spc{H}_{X_{\rm r}}\to\spc{H}_{X}\otimes\spc{H}_{M_1}$  is an isometric channel, $\map{U}_R: \rho \mapsto U_R \rho U_R^\dag$, $U_R:\spc{H}_{Z}\otimes\spc{H}_{M_1} \to \spc{H}_{Z_{\rm r}}\otimes\spc{H}_{A_{\rm r}}\otimes\spc{H}_{M_2}$ is a unitary channel, and $M_1, M_2$ are ancilla systems satisfying $d_{M_1}=d_Ad_{M_2}$. To distinguish with the original systems of $\Gamma$, subscripts are added to the recovered systems.

\begin{figure}[h]
\begin{tikzpicture}
	\begin{scope}[xscale=0.75]
		\node[virtual] (N) at (0, 0) {\phantom{$\mathcal{N}$}};
		\node[virtual] (L) at (-1.25, 0) {\phantom{$\map{S}_L$}};
		\node[virtual] (R) at (1.25, 0) {\phantom{$\map{S}_R$}};
		\node[virtual] (X) at (0, -0.75) {};
		\node[virtual] (aL) at (L|-X) {\phantom{$\map{S}_L$}};
		\node[virtual] (aR) at (R|-X) {\phantom{$\map{S}_R$}};
		\node[tensor, anchor=north, minimum height=1.25cm] (mL) at (L.north) {$\map{U}_L$};
		\node[tensor, anchor=north, minimum height=2cm] (mR) at (R.north) {$\map{U}_R$};
		\coordinate (left) at ($(L)+(-1.5,0)$);
		\coordinate (right) at ($(R)+(1.5,0)$);
		\draw (left) -- node[below, yshift=5mm]{$X_{\rm r}$} (L) -- node[below, yshift=5mm]{$X$} (N) -- node[below, yshift=5mm]{$Z$} (R) -- node[below, yshift=5mm]{$Z_{\rm r}$} (right);
		\draw (aL) -- node[below]{$M_1$} (aR) -- node[below, yshift=5mm]{$A_{\rm r}$} (aR-|right);
		\coordinate (aaR) at ($(R.east)!2!(aR.east)$);
		\draw (aaR) -- node[below, yshift=5mm]{$M_2$} (aaR-|right) node {$\times$};
	\end{scope}
\end{tikzpicture}
\caption{\label{fig:ULUR} Structure of retrodiction supermaps.}
\end{figure}

To help the analysis and presentation of the recovery supermaps, we use the following lemma that describes $\map{R}^{\map{S},\Gamma}$ with a single isometric operator:

\begin{lemma} \label{lem:rec_from_unitary}
For $\map{S}: \map{N}\mapsto \Tr_A \circ\map{N}$, assume $\Gamma$ and $\map{S}(\Gamma)$ to be full-rank and have Choi operators $C_{\Gamma}$ and $C_{\map{S}(\Gamma)}$. %
If \cref{axiom:recover_prior} is satisfied, then $\map{R}^{\map{S},\Gamma}$, represented with its corresponding CP map as defined in \cref{eq:CP_rep}, has the following form
\begin{align}
	\map{C}_{\map{R}^{\map{S},\Gamma}}(\tau) = C_\Gamma^{1/2}V\left(C_{\map{S}(\Gamma)}^{-1/2} \tau C_{\map{S}(\Gamma)}^{-1/2} \otimes \openone_{R}\right)V^\dag C_\Gamma^{1/2} \,,  \label{eq:C_R_S_Gamma} %
\end{align}
where
\begin{enumerate}
	\item $R$ is a system with $d_R \geq d_A$ and identity operator $\openone_R$, 
	\item $C_\Gamma^{1/2}$ is considered as an operator on $\spc{H}_{X_{\rm r}} \otimes \spc{H}_{Z_{\rm r}} \otimes \spc{H}_{A_{\rm r}}$, and
	\item $V$ satisfies that $VV^\dag=\openone_{X_{\rm r}Z_{\rm r}A_{\rm r}}$, namely $V^\dag: \spc{H}_{X_{\rm r}} \otimes \spc{H}_{Z_{\rm r}} \otimes \spc{H}_{A_{\rm r}} \to \spc{H}_{X} \otimes \spc{H}_{Z} \otimes \spc{H}_R$ is an isometric operator.
\end{enumerate}

Furthermore, if the rank of $\map{R}^{\map{S},\Gamma}$ is equal to $d_A$, then one has $d_R = d_A$ and $V$ is unitary.
\end{lemma}
The proof of \cref{lem:rec_from_unitary} is in \cref{app:rec_from_unitary}. %

If $d_R = d_A$ and $V$ is unitary, the retrodiction supermap in \cref{eq:C_R_S_Gamma} is analogous to a rotated Petz map in \cite{fawzi2015quantum} -- they have the same form if $V$ is a product of two unitaries commuting with $C_\Gamma$ and $C_{\map{S}(\Gamma)}\otimes\openone_{R}$, respectively.

\cref{lem:rec_from_unitary} gives a concise representation of the retrodiction supermap, which shrinks our search space for numerical optimization and makes it possible for the results to be explicitly presented.

\subsection{Choice of the prior channel}

In the following examples, we consider $X,Z,A$ being qubit systems with $d_X=d_Z=d_A=2$. The prior channel $\Gamma$ is chosen as an isometry form $X$ to $ZA$ followed by a depolarizing channel, as shown below.

\begin{center}
\begin{tikzpicture}
	\begin{scope}[xscale=0.75]
		\coordinate (X) at (0,-0.75);
		\node[virtual] (R) at (1.375, 0) {\phantom{$U_\Gamma$}};
		\node[virtual] (aR) at (R|-X) {\phantom{$U_\Gamma$}};
		\node[tensor, anchor=north, minimum height=1.25cm] (mR) at (R.north) {$U_\Gamma$};
		\node[virtual] (D) at (2.375, 0) {\phantom{$D_p$}};
		\node[virtual] (aD) at (D|-X) {\phantom{$D_p$}};
		\node[tensor, anchor=north, minimum height=1.25cm] (mD) at (D.north) {$D_p$};
		\coordinate (left) at ($(R)+(-1.25,0)$);
		\coordinate (right) at ($(R)+(2.25,0)$);
		\draw (left) -- node[above]{$X$} (R) -- (D) -- node[above]{$Z$} (right);
		\node (x) at (aR-|right) {};
		\node (zero) at (left|-aR) {$\ket0$};
		\draw (zero) -- (aR) -- (aD) -- node[below]{$A$} (x.center);
	\end{scope}
\end{tikzpicture}
\end{center}

In the diagram above, every wire is a qubit system, $U_\Gamma$ is a unitary gate, and $D_p$ is a depolarizing channel defined as
\begin{align}
	D_p(\rho) := (1-p) \rho + p \openone_{ZA} / d_{ZA}, ~0\leq p \leq 1
\end{align}
where $\openone_{ZA} / d_{ZA}$ is the maximally mixed state in $S(\spc{H}_Z \otimes \spc{H}_A)$. Picking $p>0$ ensures that both $\Gamma$ and $\map{S}(\Gamma)$ are full-rank.

\subsection{Application to error correction in cloud computing}

\begin{figure}
    \centering
    \includegraphics[width=\linewidth]{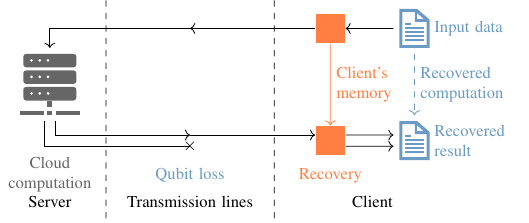}
    \caption{Supermap retrodiction applied to recover a qubit loss in cloud computing. The server receives one qubit, performs a computation procedure $\mathcal{N}$, and transmits back the result. The error is that the second qubit of the result is lost. To recover the error, the client could adopt a strategy that acts only on the erroneous result, acts only on their memory, or combines the memory and the erroneous result.}
    \label{fig:cloud_example}
\end{figure}

\red{
The result obtained here can be used as a error correction method for quantum cloud computing, illustrated in \cref{fig:cloud_example}. 
Assuming the client wants to make a computation on state $\rho$ and the server does the computation $\mathcal{N}$, the ideal errorless result is $\sigma_{\rm ideal} := \mathcal{N}(\rho)$. Here, $\rho$ is the one qubit sent from the client to the server and $\sigma_{\rm ideal}$  contains two qubits, where one of the qubits is lost during the transmission from the server back to the client, corresponding to the supermap $\map{S}$. We consider client's recovery of the error $\map{S}$ with three strategies: }

\makeatletter
\newlist{strategyenum}{enumerate}{1}
\setlist[strategyenum,1]{label={\arabic*.},
                   ref = {\arabic*}}
\crefname{strategyenumi}{Strategy}{Strategies}
\crefformat{strategyenumi}{#2\cref@strategyenumi@name~#1#3}
\crefrangeformat{strategyenumi}{#3\cref@strategyenumi@name@plural~#1#4 to #5#2#6}
\crefmultiformat{strategyenumi}{#2\cref@strategyenumi@name@plural~#1#3}{ and~#2#1#3}{, #2#1#3}{ and~#2#1#3}
\makeatother

\red{
\begin{strategyenum}
    \item \label{strategy:std_Petz} {\em Act only on the erroneous result.} Use the standard Petz map on the received qubit, with the prior chosen to be the prior channel $\Gamma$ applied to input data $\rho$. That is, apply $\map{R}^{\Tr_{A},\Gamma(\rho)}$ on the received qubit to obtain a two qubit state, resulting in recovered result  $\sigma_{\rm rec}^1 = (\map{R}^{\Tr_{A},\Gamma(\rho)} \circ \Tr_A \circ \mathcal{N} )(\rho)$. Note that this strategy requires the density matrix of $\rho$ to be known to the client, while other strategies are oblivious of $\rho$.
    \item \label{strategy:trivial} {\em Act only on the memory.} Use the trivial retrodiction supermap that ignores the server's computation and produces the prior channel $\Gamma$. Namely, store $\rho$ in the client's memory and apply $\Gamma$ on it, resulting $\sigma_{\rm rec}^2 =\Gamma(\rho)$.
    \item \label{strategy:super_retro} {\em Combine the memory and the erroneous result.} Use the retrodiction supermap $\mathcal{R}^{\mathcal{S},\Gamma}$ obtained in this section, effectively correlating the input data with the client's memory and combining the memory with the received result. The obtained recovered result is $\sigma_{\rm rec}^3 = (\mathcal{R}^{\mathcal{S},\Gamma}\circ \mathcal{S})(\mathcal{N})(\rho)$.
\end{strategyenum}
}

\red{
The recovered states in all strategies are compared with $\sigma_{\rm ideal}$. In the numerical experiments, the value of $\rho$ is chosen from the eigenbases of three Pauli matrices $\{\ket0,\ket1,(\ket0\pm\ket1)/\sqrt2, (\ket0\pm i\ket1)/\sqrt2\}$, and we will use as figure of merit the fidelity $F(\sigma_{\rm rec}, \sigma_{\rm ideal}):=\Tr\left[\sqrt{\sqrt{\sigma_{\rm rec}}\sigma_{\rm ideal}\sqrt{\sigma_{\rm rec}}}\right]^2$ averaged over all choices of $\rho$.}

\newcounter{example}
\stepcounter{example}
\subsection{Example \arabic{example}: CNOT}

For the first example, one chooses $U_\Gamma = \left(\begin{matrix}
	1 & & & \\
	& 1 & & \\
	& & & 1 \\
	& & 1 &
\end{matrix}\right)$, the CNOT gate, with system $X$ being the control system. 
The prior channel is $\Gamma(\rho) = \Gamma_{p,\mathrm{CNOT}}(\rho) := (D_p \circ \mathrm{CNOT})(\rho \otimes \ketbrasame{0})$.

We aim to find a retrodiction supermap satisfying \cref{axiom:superchannel,axiom:recover_prior}, whose Choi operator $C_{\map{R}^{\map{S},\Gamma}}$ and corresponding CP map $\map{C}_{\map{R}^{\map{S},\Gamma}}$ satisfies \cref{eq:valid_supermap,eq:C_R_S_Gamma}.
By \cref{lem:rec_from_unitary}, one can define $\map{R}^{\map{S},\Gamma}$ in terms of $V$ in \cref{eq:C_R_S_Gamma}, and under the constraint that the rank of $\map{R}^{\map{S},\Gamma}$ is no larger than $d_A=2$, $V$ is unitary.
Under the constraints \cref{eq:valid_supermap,eq:C_R_S_Gamma} and $V$ being unitary, we have found an analytical solution of $\map{R}^{\map{S},\Gamma}$ with $V$ as the following function of $p$:

\begin{align}
	V=\left(\begin{matrix} 
 1& 0& 0& 0& 0& 0& 0& 0& \\
 0& 1& 0& 0& 0& 0& 0& 0& \\
 0& 0& \cos\theta& 0& 0& -\sin\theta& 0& 0& \\
 0& 0& 0& 1& 0& 0& 0& 0& \\
 0& 0& 0& 0& 1& 0& 0& 0& \\
 0& 0& \sin\theta& 0& 0& \cos\theta& 0& 0& \\
 0& 0& 0& 0& 0& 0& 1& 0& \\
 0& 0& 0& 0& 0& 0& 0& 1& \\
\end{matrix}\right)
\end{align}
where
\begin{align}
	\sin\theta = \frac{\sqrt{8-7p} - \sqrt{p}}{2\sqrt{2-p}},~\cos\theta = \sqrt{1-\sin^2\theta}	,
\end{align}
the rows are ordered in the basis
\begin{align}
	\ket{000}_{X_{\rm r}Z_{\rm r}A_{\rm r}},\ket{001}_{X_{\rm r}Z_{\rm r}A_{\rm r}},\dots,\ket{111}_{X_{\rm r}Z_{\rm r}A_{\rm r}},
\end{align}
and the columns are ordered in the basis 
\begin{align}
	\ket{000}_{XZR},\ket{001}_{XZR},\dots,\ket{111}_{XZR}.
\end{align}

The value of $V$ is found first through numerical optimizations, then fitted to an analytical formula in terms of $p$, and last verified analytically. There may exist multiple families of solutions of $V$, and here we are presenting one of them.

The gates needed to implement the retrodiction superchannel can also be analytically described using the method in \cite{bisio2011minimal}, but they are too complicated for presentation for general $p$. In the limiting case $p\to 0$, $\sin\theta\to 1$ and $\cos\theta\to 0$. $\map{R}^{\map{S},\Gamma}$ can be implemented with the structure in \cref{fig:ULUR} with $d_{M_1}=4$ and $d_{M_2}=2$, and the isometry $U_L$ and unitary $U_R$ are defined as

\begin{align}
	U_L &= \frac{|00\rangle+|11\rangle}{\sqrt2}\langle0|+\frac{|02\rangle+|10\rangle}{\sqrt2}\langle1|\,,\label{eq:UL_CNOT}\\
	U_R &= |000\rangle\langle00|+|101\rangle\langle01|+|110\rangle\langle02|+|011\rangle\langle03| \nonumber\\
	    &~~ -|111\rangle\langle10|-|001\rangle\langle11|+|010\rangle\langle12|+|100\rangle\langle13|\,,\label{eq:UR_CNOT}
\end{align}
where the ordering of systems is $X,M_1,X_{\rm r}$ for $U_L$ and $Z_{\rm r},A_{\rm r}, M_2, Z, M_1$ for $U_R$.

The obtained retrodiction supermap does not follow the pattern mentioned in \cref{fig:select_petz_S3}. The first tooth is not measuring the system $X$ but alters its value and entangle it with the memory system $M_1$.

\begin{figure}[h]%
    \centering
    \subfigure[Both prior and true channels are $\Gamma_{p,\mathrm{CNOT}}(\rho) := (D_p \circ \mathrm{CNOT})(\rho \otimes \ketbrasame{0})$.\label{fig:compare_CNOT}]{\includegraphics[width=\linewidth, trim={2cm 0cm 1cm 0cm}]{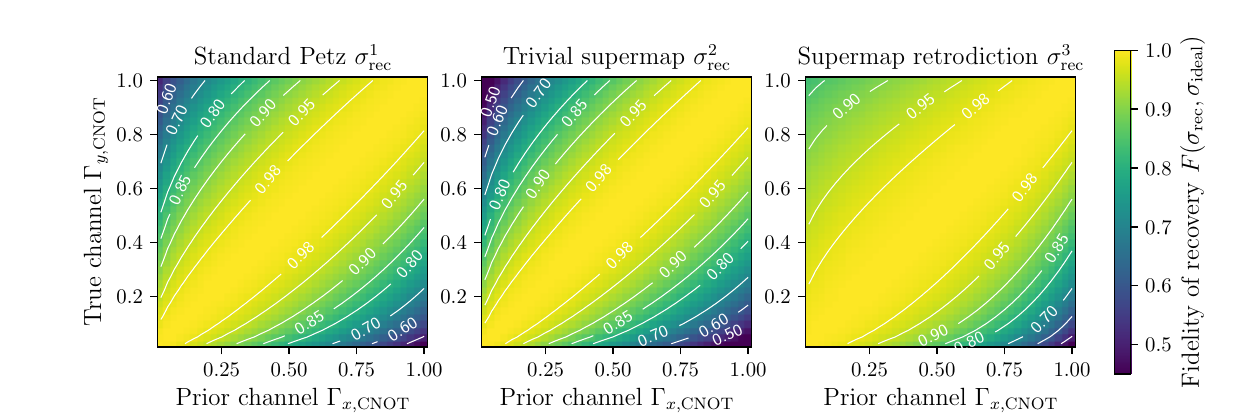}}
    
    \subfigure[Both prior and true channels are $\Gamma_{p,\mathrm{SWAP}}(\rho) := (D_p \circ \mathrm{SWAP})(\rho \otimes \ketbrasame{0})$.\label{fig:compare_SWAP}]{\includegraphics[width=\linewidth, trim={2cm 0cm 1cm 0cm}]{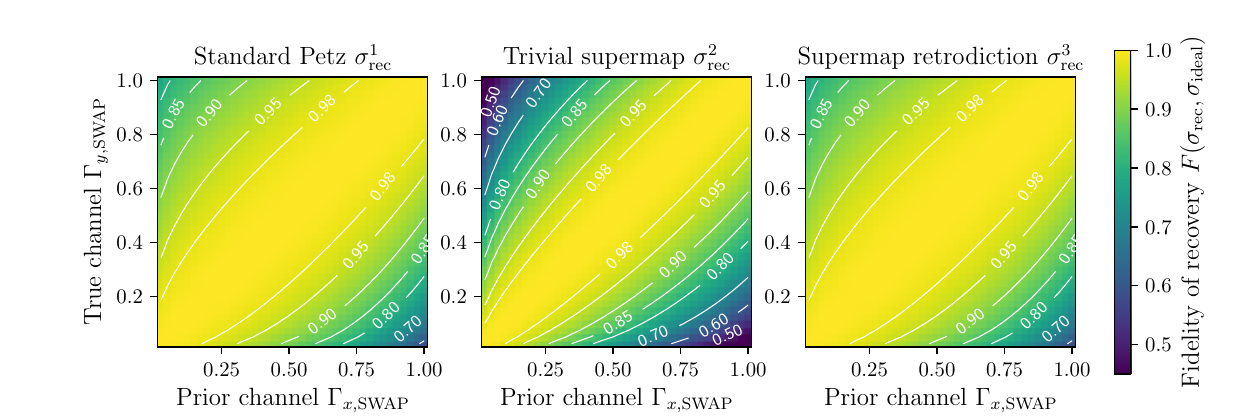}}
    
    \subfigure[Both prior and true channels are $\Gamma_{p,\mathrm{id}}(\rho) := D_p(\rho \otimes \ketbrasame{0})$.\label{fig:compare_identity}]{\includegraphics[width=\linewidth, trim={2cm 0cm 1cm 0cm}]{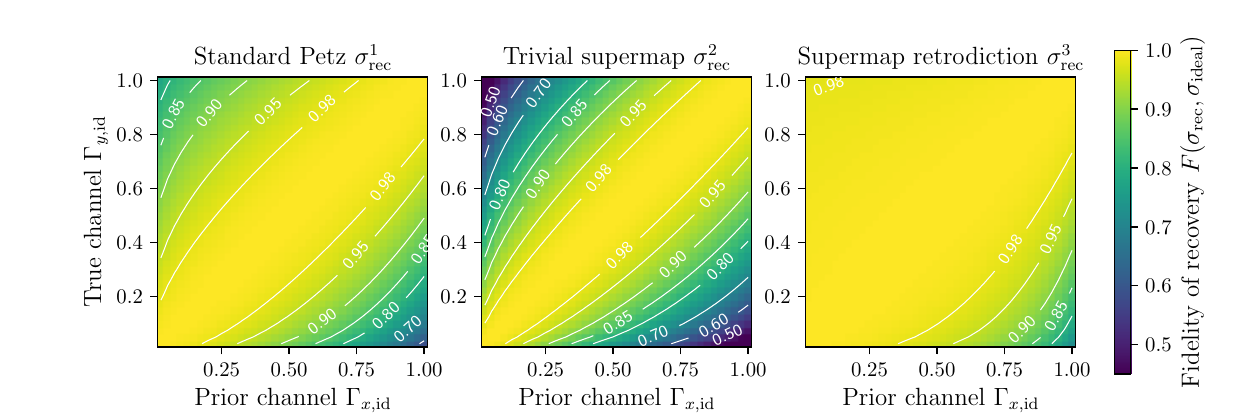}}

    \caption{\red{Comparison between strategies used for error correction in quantum cloud computing. In each figure, the horizontal axis denotes the parameter in the prior channel, and the vertical axis denotes the parameter in the real channel, which may be different from the prior one. Along the $x=y$ diagonals of figures, the real channel equals to the prior channel, and all strategies reach fidelity one. 
    In all cases, the supermap retrodiction obtained in this section performs no worse than the other strategies. }}%
    \label{fig:compare}
\end{figure}

\red{
Now, we use the obtained retrodiction supermap for error correction in cloud computing, and  compare it with other strategies, with numerical results shown in \cref{fig:compare_CNOT}. The horizontal axis is the parameter in the prior channel $\Gamma_{x,\mathrm{CNOT}}$, while the vertical axis is the parameter in the true channel $\Gamma_{y,\mathrm{CNOT}}$.
Along the $x=y$ diagonal, the true channel equals to the prior channel, and all strategies recover the ideal result with fidelity one.
In other regions, the retrodiction supermap obtained in this section performs better than the other two methods, even though \cref{strategy:std_Petz} uses additional information about $\rho$ in constructing the recovery map.
}

\stepcounter{example}
\subsection{Example \arabic{example}: SWAP}

Here $U_\Gamma$ is chosen as $\left(\begin{matrix}
	1 & & & \\
	& & 1 & \\
	& 1 & & \\
	& & & 1
\end{matrix}\right)$, the SWAP gate, and $\Gamma(\rho) = \Gamma_{p,\mathrm{SWAP}}(\rho) := (D_p \circ \mathrm{SWAP})(\rho \otimes \ketbrasame{0})$.
Similar to the previous example, the retrodiction supermap $\map{R}^{\map{S},\Gamma}$ can be defined in terms of $V$ in \cref{eq:C_R_S_Gamma} as
\begin{align}
	\displaystyle V = \left(\begin{matrix}
0 & 0 & 0 & 0 & -1 & 0 & 0 & 0\\
0 & 0 & 0 & 0 & 0 & 1 & 0 & 0\\
0 & 0 & 0 & 0 & 0 & 0 & 0 & 1\\
0 & 0 & 0 & \cos\theta & 0 & 0 & -\sin\theta & 0\\
1 & 0 & 0 & 0 & 0 & 0 & 0 & 0\\
0 & 1 & 0 & 0 & 0 & 0 & 0 & 0\\
0 & 0 & 0 & \sin\theta & 0 & 0 & \cos\theta & 0\\
0 & 0 & 1 & 0 & 0 & 0 & 0 & 0
\end{matrix}\right)
\end{align}
\begin{align}
	\cos\theta = \frac{\sqrt{8-7p} - \sqrt{p}}{2\sqrt{2-p}}, ~ \sin\theta = \sqrt{1-\cos^2\theta}
\end{align}

In the limiting case $p\to 0$, $\map{R}^{\map{S},\Gamma}$ can be implemented with the structure in \cref{fig:ULUR} with $d_{M_1}=4$ and $d_{M_2}=2$, and the isometry $U_L$ and unitary $U_R$ are defined as
\begin{align}
	U_L &= \frac{|00\rangle+|11\rangle}{\sqrt2}\langle0|+\frac{|02\rangle+|13\rangle}{\sqrt2}\langle1| \,,\\
	U_R &= |000\rangle\langle00|+|001\rangle\langle01|+|010\rangle\langle02|+|011\rangle\langle03|\nonumber\\
	&~~ +|110\rangle\langle10|+|100\rangle\langle11|-|111\rangle\langle12|-|101\rangle\langle13| \,,
\end{align}
with the same ordering of systems as \cref{eq:UL_CNOT,eq:UR_CNOT}.

\red{A comparison of this retrodiction supermap with other strategies for error correction in cloud computing is shown in \cref{fig:compare_SWAP}. 
Again, we observe that the retrodiction supermap obtained in this section performs no worse than the other two methods. The performances of \cref{strategy:std_Petz,strategy:super_retro} are equal up to numerical precision ($10^{-12}$).}

\stepcounter{example}
\subsection{Example \arabic{example}: Identity} \label{ss:identity}

Here, one chooses $U_\Gamma = \openone$, and $\Gamma(\rho) =\Gamma_{p,\mathrm{id}}(\rho) :=D_p(\rho \otimes \ketbrasame{0})$. The retrodiction supermap $\map{R}^{\map{S},\Gamma}$ written in terms of $V$ in \cref{eq:C_R_S_Gamma} is

\begin{align}
V = \left(\begin{matrix} 
 0& 1& 0& 0& 0& 0& 0& 0& \\
 0& 0& 0& \sin\theta& \cos\theta& 0& 0& 0& \\
 1& 0& 0& 0& 0& 0& 0& 0& \\
 0& 0& 1& 0& 0& 0& 0& 0& \\
 0& 0& 0& 0& 0& 0& 1& 0& \\
 0& 0& 0& 0& 0& 1& 0& 0& \\
 0& 0& 0& 0& 0& 0& 0& 1& \\
 0& 0& 0& \cos\theta& -\sin\theta& 0& 0& 0& \\
\end{matrix}\right)
\end{align}
\begin{align}
	\sin\theta = \sqrt{\frac{\sqrt{p}+\sqrt{8-7p}}{2\sqrt{4-3p}}}, ~ \cos\theta = \sqrt{1-\sin^2\theta}
\end{align}

In the limiting case $p\to 0$, $\map{R}^{\map{S},\Gamma}$ can be implemented with $d_{M_1}=4$, $d_{M_2}=2$, and $U_L$ and $U_R$ defined in the following, with the same ordering of systems as \cref{eq:UL_CNOT,eq:UR_CNOT}.

\begin{align}
	&U_L = \left(\frac{\sqrt{2+\sqrt2}}{2}|00\rangle+\frac{\sqrt{2-\sqrt2}}{2}|11\rangle\right)\langle0| \nonumber\\
	 &\qquad+\left(\frac{\sqrt{2-\sqrt2}}{2}|02\rangle+\frac{\sqrt{2+\sqrt2}}{2}|13\rangle\right)\langle1| \,,\\
	& U_R  = \ketbra{011}{01} + \ketbra{110}{12}\nonumber \\
	&+ \frac{\sqrt{2-\sqrt2}}{2}\left(\ketbra{101}{00} + \ketbra{000}{11} +\ketbra{001}{13} + \ketbra{100}{02}\right)\nonumber \\
	&+ \frac{\sqrt{2+\sqrt2}}{2}\left(\ketbra{000}{00} -\ketbra{101}{11} +\ketbra{100}{13} - \ketbra{001}{02}\right)\nonumber \\
	&+ \sqrt{\frac{2}{2+\sqrt2}}\left(\ketbra{010}{03} + \ketbra{111}{10}\right) \nonumber \\
	&+ \sqrt{\frac{\sqrt2}{2+\sqrt2}}\left(\ketbra{010}{10} - \ketbra{111}{03}\right) \,.
\end{align}

\red{A comparison of this retrodiction supermap with other strategies for error correction in cloud computing is shown in \cref{fig:compare_identity}. 
In this case, \cref{strategy:super_retro} reaches high fidelity for most choices of $x$ and $y$. }%

\begin{figure}[h]%
    \centering
    \includegraphics[width=\linewidth, trim={2cm 0cm 1cm 0cm}]{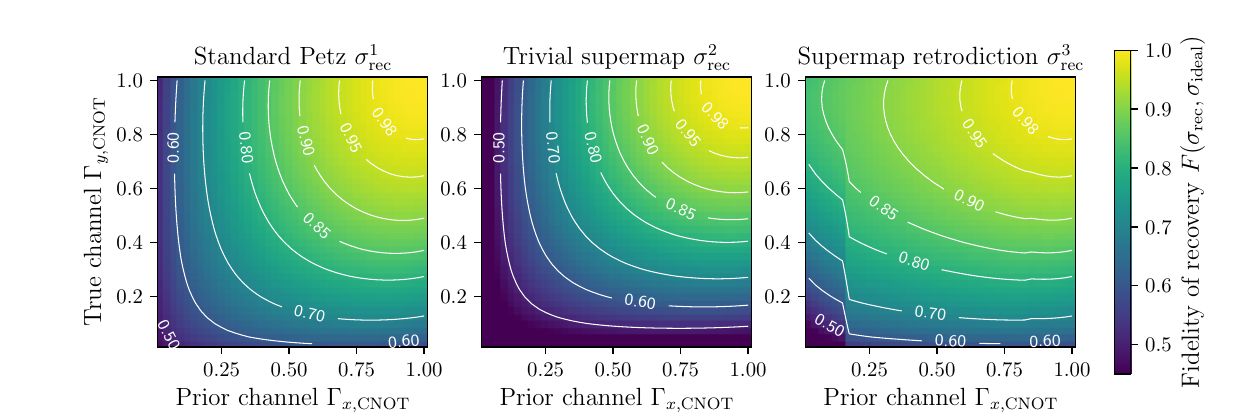}
    \caption{\red{Comparison between error-correction strategies for quantum cloud computing. The prior channels are $\Gamma_{x,\mathrm{CNOT}}$ while the true channels are $\Gamma_{y,\mathrm{SWAP}}$.}}
    \label{fig:compare_CNOT_SWAP}
\end{figure}

\red{Last, we consider the case where the prior and true channels are from different classes, $\Gamma_{x,\mathrm{CNOT}}$ and $\Gamma_{y,\mathrm{SWAP}}$, and check whether supermap retrodiction is able to correct the error. The results are shown in \cref{fig:compare_CNOT_SWAP}. The performance of \cref{strategy:super_retro} is better than the other strategies for most values of $x,y$, with the exception that \cref{strategy:std_Petz} is slightly better for some regions such as the lower left corner in the figure. This may indicate that using a prior that is too confident (close to an isometry whose Choi operator is rank-one) may give lower performance if the reality is very different. Yet, \cref{strategy:super_retro} has wider applicability than \cref{strategy:std_Petz} since the client could perform error correction for input data of which they do not have any information.}

\section{Discussion} \label{sec:discussion}

Our work gives a framework and a partial solution to the Bayesian retrodiction of quantum superchannels. For general cases, we have found analytical solutions to a few families of examples. The solutions are similar to a rotated Petz map, with rotations for which we do not have a general recipe.

The retrodiction of quantum superchannels, which is also the update rule for beliefs of quantum channels, is a basic component of quantum Bayesian networks \cite{tucci1995quantum,warmuth2005bayes,tucci2007factorization,leifer2008quantum,barrett2019quantum}. A classical Bayesian network \cite{pearl} is a connection of random variables with conditional probabilities. It is a machine learning model where the connections can be updated according to observations, and later used for making predictions.
The quantum Bayesian network is a connection of quantum systems with quantum channels, where the channels can be updated according to observations and used for predictions. 
The Bayesian method may not be the optimal solution for certain tasks (for example, the Bayes' rule is not optimal for state retrieval \cite{surace2022state}), but will hopefully be more consistent and scalable than numerical optimizations. %

Compared with other proposals to update beliefs in quantum Bayesian networks \cite{warmuth2005bayes,leifer2008quantum}, our proposal of supermap retrodiction is both conceptually consistent with Bayes' rule and operationally realizable with a deterministic quantum circuit. This has the following benefits.%

First, the retrodiction supermap can be used to recover errors of quantum operations \red{as exemplified in \cref{sec:examples}}. %
The error model is a supermap capable of characterizing errors on the input, the output and unwanted side channel between them. This model is particularly suitable for accessing a remote process, such as cloud computing \cite{arrighi2006blind,broadbent2009universal,morimae2013blind} and quantum illumination \cite{lloyd2008enhanced}, where errors may occur at the transmission in both directions.

Second, the retrodiction can be applied to subsystems of a quantum process, and the quantum nature of our proposal makes it possible to preserve the entanglement between the subsystem of interest and its complement. In contrast, collecting the observations as classical data and making conceptual belief updates necessarily destroy entanglement.

Although we have found solutions of retrodiction supermaps for subclasses of superchannels and priors, a universal recipe is yet to be found. %
It remains unknown whether the aforementioned axioms can be all satisfied by a universal recipe. \red{Specifically, it is unknown whether the involutive property is still applicable to supermap retrodiction.} It is possible that some of them have to be compromised, for example, lifting \cref{axiom:superchannel} to allow for probabilistic supermaps \cite{chiribella2008transforming} or virtual supermaps (weighted difference between two superchannels) \cite{zhu2024reversing}. They are still physical in the sense that they can be simulated with deterministic circuits and classical post-processing at the cost of more experimental repetitions.

\red{The main idea of this work is placing a quantum channel at the position of a state in the conventional quantum Bayes' rule. This may not be the end of the story, since we may consider scenarios where the client has more complicated priors, such as  prior beliefs on both their input data and the server's computation. In this case, a supermap can be considered mapping a state plus a channel to another state. This can be included in a more general setting, that is, beliefs of higher-order quantum operations like quantum combs \cite{chiribella2008quantum,chiribella2009theoretical} or even quantum processes without a definite causal structure \cite{chiribella2013quantum}. This work, dealing with a relatively simple yet non-trivial case in this direction, may lead to belief updates that incorporates with various forms of prior knowledge, %
with wide applications in quantum communication, quantum metrology and quantum machine learning.}

\section*{Code availability}
The code used to perform the numerical experiments is available at \url{https://github.com/bg95/supermap_retrodiction}.

\section*{Acknowledgments}
We thank Valerio Scarani for helpful discussions and suggestions. 

This research is supported by the National Research Foundation, Singapore and A*STAR under its CQT Bridging Grant; by the Ministry of Education, Singapore, under the Tier 2 grant ``Bayesian approach to irreversibility'' (Grant No.~MOE-T2EP50123-0002); and by the Start-up Fund (Grant No.~G0101000274) from The Hong Kong University of Science and Technology (Guangzhou).

\bibliography{retro_supermap}

\appendix

\section{Example of Petz map violating \cref{axiom:superchannel}}
\label{app:Petz_not_superchannel}

Consider the superchannel $\map{S}: \map{N}\mapsto\Tr_A\circ\map{N}$ and the prior channel $\Gamma$ with the structure in the following figure
\begin{center}
    \begin{tikzpicture}
    \node[virtual] (R) at (1.375, 0) {\phantom{$\Gamma$}};
    \node[virtual] (aR) at (R|-W) {\phantom{$\Gamma$}};
    \node[tensor, anchor=north, minimum height=1.25cm] (mR) at (R.north) {$\Gamma$};
    \coordinate (left) at ($(R)+(-1.25,0)$);
    \coordinate (right) at ($(R)+(1.25,0)$);
    \coordinate (right2) at ($(right)+(0.5,0)$);
    \coordinate (below) at (0, -1.4);
    \draw (left) -- node[above]{$W$} (R) -- node[above]{$Z$} (right) -- (right2);
    \node (x) at (aR-|right) {$\times$};
    \draw (aR) -- node[below]{$A$} (x.center);
    \draw[dashed] ($(R.north east-|right)+(-0.2,0.2)$) rectangle ($(aR.south east-|right)+(0.2,-0.2)$);
    \node (Wprime) at (x|-below) {$\map{S}$};
\end{tikzpicture}
\end{center}
where $d_W=d_Z=d_A=2$. $\Gamma$ is defined as
\begin{align}
    \Gamma(\rho) := \frac12(\rho \otimes \ketbrasame{0}) + \frac12 \frac{\openone_{ZA}}{4},
\end{align}
which mixes $\rho\otimes\ketbrasame{0}$ with the maximally mixed state $\openone_{ZA}/4$. This is indeed the example in \cref{ss:identity} with $p=1/2$. With the Choi--\Jami{} isomorphism,  $\map{S}$, $\Gamma$ and $\map{S}(\Gamma)$ are mapped to
\begin{align}
    \map{C}_\map{S} &= \Tr_A \\
    C_{\Gamma} &= \frac12\kketbbrasame{\openone}_{WZ}\otimes\ketbrasame{0}_A + \frac{\openone_{WZA}}{8} \\
    C_{\map{S}(\Gamma)} &= \map{C}_\map{S}(C_{\Gamma})=\kketbbrasame{\openone}_{WZ}+ \frac{\openone_{WZ}}{4} 
\end{align}
where $\kket{\openone}_{WZ}:=\ket{00}_{WZ}+\ket{11}_{WZ}$. The components for the Petz map can be calculated as
\begin{align}
    \map{C}_\map{S}^\dag(M) &= M \otimes \openone_A\\
    \sqrt{C_{\Gamma}} &= \frac1{\sqrt8}\kketbbrasame{\openone}_{WZ}\otimes\ketbrasame{0}_A + \frac{\openone_{WZA}}{\sqrt8} \\
    C_{\map{S}(\Gamma)}^{-1/2} &= \left(\frac{1}{\sqrt5}-1\right)\kketbbrasame{\openone}_{WZ}+ 2\openone_{WZ}
\end{align}
The expression of the Petz map in \cref{eq:Petz} gives the following retrodiction supermap
\begin{align}
    \map{C}_{\map{R}^{\map{S},\Gamma}}(M) = \sqrt{C_{\Gamma}}\left( C_{\map{S}(\Gamma)}^{-1/2}
 M C_{\map{S}(\Gamma)}^{-1/2} \otimes \openone_A \right)\sqrt{C_{\Gamma}}
\end{align}
Denoting the recovered systems as $Z_{\rm r}, A_{\rm r}, W_{\rm r}$, the Choi matrix of $\map{C}_{\map{R}^{\map{S},\Gamma}}$ (or equivalently $\map{R}^{\map{S},\Gamma}$) is
\begin{align}
    & C_{\map{R}^{\map{S},\Gamma}} = (\openone_{WZ}\otimes K) \nonumber\\
    & \times \left(\kketbbrasame{\openone}_{WW_{\rm r}}\otimes\kketbbrasame{\openone}_{ZZ_{\rm r}} \otimes \openone_{A_{\rm r}}\right) (\openone_{WZ}\otimes K^\dag) 
\end{align}
where
\begin{align}
     &K := \frac{\openone_{W_{\rm r}Z_{\rm r}A_{\rm r}}}{\sqrt2} \nonumber\\
     &+\kketbbrasame{\openone}_{W_{\rm r}Z_{\rm r}}\otimes\left(\frac{3\sqrt5 - 5}{10\sqrt2}\ketbrasame{0}_{A_{\rm r}} + \frac{\sqrt5 - 5}{10\sqrt2} \ketbrasame{1}_{A_{\rm r}}\right)
\end{align}

Notice $\map{R}^{\map{S},\Gamma}$ is a supermap from $\CP(\spc{H}_{W}, \spc{H}_{Z})$ to $\CP(\spc{H}_{W_{\rm r}}, \spc{H}_{Z_{\rm r}}\otimes\spc{H}_{A_{\rm r}})$, where $\mathrm{CP}$ denotes the set of completely positive maps. The first equation of the criteria \cref{eq:valid_supermap} reads
\begin{align} \label{eq:example_criteria}
    \Tr_{Z_{\rm r}A_{\rm r}}\left[C_{\map{R}^{\map{S},\Gamma}}\right] = \Tr_{ZZ_{\rm r}A_{\rm r}}\left[C_{\map{R}^{\map{S},\Gamma}}\right] / d_Z \otimes \openone_Z
\end{align}
The left hand side, written in the basis $\ket{000}_{WW_{\rm r}Z},\ket{001}_{WW_{\rm r}Z},\dots,\ket{111}_{WW_{\rm r}Z}$, equals
\begin{widetext}
    \begin{align}
    \Tr_{Z_{\rm r}A_{\rm r}}\left[C_{\map{R}^{\map{S},\Gamma}}\right] = 
    \left(\begin{matrix}\frac{\sqrt{5}}{5} + \frac{1}{2} & 0 & 0 & 0 & 0 & 0 & \frac{\sqrt{5}}{5} + \frac{1}{2} & 0\\0 & 1 & 0 & 0 & - \frac{1}{2} + \frac{\sqrt{5}}{5} & 0 & 0 & \frac{\sqrt{5}}{5} + \frac{1}{2}\\0 & 0 & 0 & 0 & 0 & 0 & 0 & 0\\0 & 0 & 0 & \frac{1}{2} - \frac{\sqrt{5}}{5} & 0 & 0 & - \frac{1}{2} + \frac{\sqrt{5}}{5} & 0\\0 & - \frac{1}{2} + \frac{\sqrt{5}}{5} & 0 & 0 & \frac{1}{2} - \frac{\sqrt{5}}{5} & 0 & 0 & 0\\0 & 0 & 0 & 0 & 0 & 0 & 0 & 0\\\frac{\sqrt{5}}{5} + \frac{1}{2} & 0 & 0 & - \frac{1}{2} + \frac{\sqrt{5}}{5} & 0 & 0 & 1 & 0\\0 & \frac{\sqrt{5}}{5} + \frac{1}{2} & 0 & 0 & 0 & 0 & 0 & \frac{\sqrt{5}}{5} + \frac{1}{2}\end{matrix}\right)
\end{align}
\end{widetext}
This is not in a tensor product form with $\openone_Z$ , and therefore \cref{eq:example_criteria} does not hold. Thus $\map{R}^{\map{S},\Gamma}$ is not a superchannel.

\section{Proof of \cref{lem:rec_from_unitary}} \label{app:rec_from_unitary}
Let $\map{R} := \map{R}^{\map{S},\Gamma}$ for simplicity.
We use the correspondence in \cref{eq:CP_rep} and denote the CP map corresponding to $\map{R}$ as $\map{C}_{\map{R}} \in \mathrm{CP}(\spc{H}_W\otimes\spc{H}_Z\otimes\spc{H}_A,\spc{H}_{W_{\rm r}}\otimes\spc{H}_{Z_{\rm r}})$, $\map{C}_{\map{R}}:C_{\map{N}} \mapsto C_{\map{R}(\map{N})}$.

The Kraus rank of $\map{C}_{\map{R}}$ is equal to the rank of $\map{R}$, since their Choi operators are equal up to reordering of systems. Let the rank of $\map{C}_{\map{R}}$ be $d_R$, and we can write the Kraus decomposition of $\map{C}_{\map{R}}$ as 
\begin{align} \label{eq:Kraus_C}
	\map{C}_{\map{R}}(C_{\map{N}})=C_{\map{R}(\map{N})} = \sum_{k=1}^{d_R} K_k C_{\map{N}} K_k^\dag \,.
\end{align}
Defining $K := \sum_{k=1}^{d_R} K_k \otimes \bra{k}$, where $\{\ket{k}\}$ is a basis of Hilbert space $\spc{H}_R$ with dimension $d_R$, we can rewrite \cref{eq:Kraus_C} as
\begin{align}
	C_{\map{R}(\map{N})} = K (C_{\map{N}} \otimes \openone_R) K^\dag \,,
\end{align}
where $\openone_R$ is the identity operator on $\spc{H}_R$.
Defining $V:= C_\Gamma^{-1/2} K (C_{\map{S}(\Gamma)}^{1/2} \otimes\openone_R)$, this equation becomes
\begin{align} \label{eq:C_R_S_Gamma_channel}
	C_{\map{R}(\map{N})} = C_\Gamma^{1/2} V (C_{\map{S}(\Gamma)}^{-1/2} C_{\map{N}} C_{\map{S}(\Gamma)}^{-1/2} \otimes \openone_R) V^\dag C_\Gamma^{1/2}\,.
\end{align}
This is equivalent to \cref{eq:C_R_S_Gamma}.

Now, we show that $V^\dag$ is an isometry. Substituting $\map{N}$ with $\map{S}(\Gamma)$ in \cref{eq:C_R_S_Gamma_channel}, we get
\begin{align}
	C_{\map{R}(\map{S}(\Gamma))} &= C_\Gamma^{1/2} V (C_{\map{S}(\Gamma)}^{-1/2} C_{\map{S}(\Gamma)} C_{\map{S}(\Gamma)}^{-1/2} \otimes \openone_R) V^\dag C_\Gamma^{1/2} \nonumber\\
	 &= C_\Gamma^{1/2} VV^\dag C_\Gamma^{1/2}\,.
\end{align}
By \cref{axiom:recover_prior}, $\map{R}(\map{S}(\Gamma)) = \Gamma$, thus
\begin{align}
	C_\Gamma &= C_\Gamma^{1/2} VV^\dag C_\Gamma^{1/2} \\
	\openone_{W_{\rm r}Z_{\rm r}A_{\rm r}} &= VV^\dag
\end{align}
and $V^\dag$ is an isometry. $V^\dag$ being an isometry automatically indicates $d_R \geq d_A$. If the rank of $\map{R}$ is equal to $d_A$, then $d_R = d_A$ and $V$ is unitary.

\end{document}

%% file: header.tex
\usepackage{amssymb,enumerate}
\usepackage{graphicx}
\usepackage{graphics}
\usepackage{amsmath}
\usepackage{amsthm,bbm}
\usepackage{color}
\usepackage{dsfont}
\usepackage{hyperref}

\usepackage[capitalise,nameinlink]{cleveref}

\usepackage{qcircuit}

\usepackage{graphicx}

\usepackage{xcolor}
\hypersetup{
    colorlinks,
    linkcolor={red!50!black},
    citecolor={blue!50!black},
    urlcolor={blue!80!black}
}

\usepackage{xfrac}

\usepackage{amsmath}
\usepackage{tikz-cd}
\usepackage{empheq}

\usepackage{enumitem}

\usepackage{graphicx}%

\usepackage{hyperref}
\usepackage[capitalise]{cleveref}
\usepackage{tikz}
\usepackage{mathtools}

\newcommand{\C}{\mathbb{C}}

\newcommand{\Tr}{\operatorname{Tr}}

\newcommand{\<}{\langle}
\renewcommand{\>}{\rangle}

\newcommand{\spc}[1]{\mathcal{#1}}
\newcommand{\map}[1]{\mathcal{#1}}

\newtheorem{thm}{Theorem}

\usepackage{lmodern}
\usepackage{tcolorbox}

\usepackage{amsfonts}
\usepackage{graphicx,graphics,epsfig,times,bm,bbm,amssymb,amsmath,amsfonts,mathrsfs}
\usepackage[normalem]{ulem}
\usepackage{setspace}
\usepackage{subfigure}
\usepackage{dsfont}
\usepackage{braket}
\usepackage{upgreek}
\usepackage{tikz}
\usepackage{natbib}
\usepackage{chngcntr}

\newtheorem{theorem}{Theorem}

\newtheorem{lemma}[theorem]{Lemma}

\makeatletter 
\renewcommand\onecolumngrid{%
  \do@columngrid{one}{\@ne}%
  \def\set@footnotewidth{\onecolumngrid}%
  \def\footnoterule{\kern-6pt\hrule width 1.5in\kern6pt}%
}
\makeatother

\newcommand{\red}[1]{\textcolor{red}{#1}}

\newcommand{\bes} {\begin{subequations}}
\newcommand{\ees} {\end{subequations}}
\newcommand{\bea} {\begin{eqnarray}}
\newcommand{\eea} {\end{eqnarray}}
\newcommand{\be} {\begin{equation}}
\newcommand{\ee} {\end{equation}}

\def\>{\rangle}
\def\<{\langle}
\def\Tr{\mathrm{Tr}}

\newcommand{\ketbra}[2]{|{#1}\rangle\langle{#2}|}
\newcommand{\ketbrasame}[1]{\ketbra{#1}{#1}}
\newcommand{\kket}[1]{|#1\rangle\!\rangle}
\newcommand{\bbra}[1]{\langle\!\langle#1|}
\newcommand{\kketbbrasame}[1]{\kket{#1}\bbra{#1}}

\newcommand{\ignore}[1]{}

\newcommand{\Jami}{Jamio\l{}kowski}

%% file: tikzsetting.tex
\usetikzlibrary{decorations.pathreplacing, positioning, shapes.misc}
\tikzset{tensor/.style={rectangle,color=black,draw=black,fill=white,
                    inner sep=1pt,minimum size=5mm}}
\tikzset{tensor2h/.style={rectangle,color=black,draw=black,fill=white,
                    inner sep=1pt,minimum width=5mm,minimum height=10mm}}
\tikzset{tensor3h/.style={rectangle,color=black,draw=black,fill=white,
                    inner sep=1pt,minimum width=5mm,minimum height=15mm}}
\tikzset{parameter/.style={rectangle,color=black,draw=black,fill=black!10,thick,
                    inner sep=1pt,minimum size=5mm}}
\tikzset{virtual/.style={rectangle,inner sep=1pt,minimum size=5mm}}
\tikzset{prepare/.style={rounded rectangle, rounded rectangle east arc=none,color=black,draw=black,fill=white,inner sep=1pt,minimum size=5mm}}
\tikzset{measure/.style={rounded rectangle, rounded rectangle west arc=none,color=black,draw=black,fill=white,inner sep=1pt,minimum size=5mm}}
\tikzset{
    triple3/.style args={[#1] in [#2] in [#3]}{
        #1,preaction={preaction={draw,#3},draw,#2}
    }
}
\tikzset{triple/.style={triple3={[line width=0.125mm,black] in [line width=2mm,white] in [line width=2.25mm,black]}}}
\tikzset{thick triple/.style={triple3={[line width=0.25mm,black] in [line width=1.75mm,white] in [line width=2.25mm,black]}}}

\tikzset{measurement/.pic={\draw (55:0.5) arc (55:125:0.5); \draw (80:0.25) -- (80:0.6);}}
\tikzset{ground/.pic={\draw[thick] (-0.15,0) -- (0.15,0);\draw[thick] (-0.10,-0.05) -- (0.10,-0.05);\draw[thick] (-0.05,-0.1) -- (0.05,-0.1);}}
\tikzset{cross/.pic={\draw (-0.1,-0.1) -- (0.1,0.1);\draw (-0.1,0.1) -- (0.1,-0.1);}}
\tikzset{x/.pic={\draw (-0.1,-0.1) -- (0.1,0.1);\draw (-0.1,0.1) -- (0.1,-0.1);}}

\tikzset{c1/.pic={\fill (0, 0) circle [radius=0.1];}}
\tikzset{c0/.pic={\fill[white] (0, 0) circle [radius=0.1]; \draw (0, 0) circle [radius=0.1];}}
\tikzset{c01/.pic={
    \fill[white] (0, 0) circle [radius=0.1];
    \draw[fill] (0,0) -- (-135:0.1) arc (-135:45:0.1) -- cycle;
    \draw (0,0) circle [radius=0.1];
    }
}

\tikzset{tooth down/.style={
        draw=none, opacity=0, text opacity=1,
        append after command={
            ([shift={(-1\pgflinewidth,+0.5\pgflinewidth)}]\tikzlastnode.north east) 
        edge([shift={(-1\pgflinewidth,-0.5\pgflinewidth)}]\tikzlastnode.south east) 
            (\tikzlastnode.south east)
        edge(\tikzlastnode.south west)            
            ([shift={(0,-0.5\pgflinewidth)}]\tikzlastnode.south west)
        edge([shift={(0,+0.5\pgflinewidth)}]\tikzlastnode.north west)
        }
    }
}

\tikzset{omit right/.style={
        draw=none,
        append after command={
            [shorten <= -0.5\pgflinewidth]
            ([shift={(-1.5\pgflinewidth,-0.5\pgflinewidth)}]\tikzlastnode.north east)
        edge([shift={( 0.5\pgflinewidth,-0.5\pgflinewidth)}]\tikzlastnode.north west) 
            ([shift={( 0.5\pgflinewidth,-0.5\pgflinewidth)}]\tikzlastnode.north west)
        edge([shift={( 0.5\pgflinewidth,+0.5\pgflinewidth)}]\tikzlastnode.south west)            
            ([shift={( 0.5\pgflinewidth,+0.5\pgflinewidth)}]\tikzlastnode.south west)
        edge([shift={(-1.0\pgflinewidth,+0.5\pgflinewidth)}]\tikzlastnode.south east)
        }
    }
}

\tikzset{tooth/.style={
        draw=none,
        append after command={
            [shorten <= -0.5\pgflinewidth]
            ([shift={(-0.5\pgflinewidth,+1.5\pgflinewidth)}]\tikzlastnode.south east)
        edge([shift={(-0.5\pgflinewidth,-0.5\pgflinewidth)}]\tikzlastnode.north east) 
            ([shift={(-0.5\pgflinewidth,-0.5\pgflinewidth)}]\tikzlastnode.north east)
        edge([shift={( 0.5\pgflinewidth,-0.5\pgflinewidth)}]\tikzlastnode.north west)            
            ([shift={( 0.5\pgflinewidth,-0.5\pgflinewidth)}]\tikzlastnode.north west)
        edge([shift={(+0.5\pgflinewidth,+1.0\pgflinewidth)}]\tikzlastnode.south west)
        }
    }
}

\tikzset{tooth down/.style={
        draw=none,
        append after command={
            [shorten <= -0.5\pgflinewidth]
            ([shift={(-0.5\pgflinewidth,-1.5\pgflinewidth)}]\tikzlastnode.north east)
        edge([shift={(-0.5\pgflinewidth,+0.5\pgflinewidth)}]\tikzlastnode.south east) 
            ([shift={(-0.5\pgflinewidth,+0.5\pgflinewidth)}]\tikzlastnode.south east)
        edge([shift={( 0.5\pgflinewidth,+0.5\pgflinewidth)}]\tikzlastnode.south west)            
            ([shift={( 0.5\pgflinewidth,+0.5\pgflinewidth)}]\tikzlastnode.south west)
        edge([shift={(+0.5\pgflinewidth,-1.0\pgflinewidth)}]\tikzlastnode.north west)
        }
    }
}

\makeatletter

\pgfkeys{/pgf/.cd,
  rectangle corner radius north west/.initial=10pt,
  rectangle corner radius north east/.initial=10pt,
  rectangle corner radius south west/.initial=10pt,
  rectangle corner radius south east/.initial=10pt
}
\newif\ifpgf@rectanglewrc@donecorner@
\def\pgf@rectanglewithroundedcorners@docorner#1#2#3#4#5{%
  \edef\pgf@marshal{%
    \noexpand\pgfintersectionofpaths
      {%
        \noexpand\pgfpathmoveto{\noexpand\pgfpoint{\the\pgf@xa}{\the\pgf@ya}}%
        \noexpand\pgfpathlineto{\noexpand\pgfpoint{\the\pgf@x}{\the\pgf@y}}%
      }%
      {%
        \noexpand\pgfpathmoveto{\noexpand\pgfpointadd
          {\noexpand\pgfpoint{\the\pgf@xc}{\the\pgf@yc}}%
          {\noexpand\pgfpoint{#1}{#2}}}%
        \noexpand\pgfpatharc{#3}{#4}{#5}%
      }%
    }%
  \pgf@process{\pgf@marshal\pgfpointintersectionsolution{1}}%
  \pgf@process{\pgftransforminvert\pgfpointtransformed{}}%
  \pgf@rectanglewrc@donecorner@true
}
\pgfdeclareshape{rectangle with rounded corners}
{
  \inheritsavedanchors[from=rectangle] %
  \inheritanchor[from=rectangle]{north}
  \inheritanchor[from=rectangle]{north west}
  \inheritanchor[from=rectangle]{north east}
  \inheritanchor[from=rectangle]{center}
  \inheritanchor[from=rectangle]{west}
  \inheritanchor[from=rectangle]{east}
  \inheritanchor[from=rectangle]{mid}
  \inheritanchor[from=rectangle]{mid west}
  \inheritanchor[from=rectangle]{mid east}
  \inheritanchor[from=rectangle]{base}
  \inheritanchor[from=rectangle]{base west}
  \inheritanchor[from=rectangle]{base east}
  \inheritanchor[from=rectangle]{south}
  \inheritanchor[from=rectangle]{south west}
  \inheritanchor[from=rectangle]{south east}

  \savedmacro\cornerradiusnw{%
    \edef\cornerradiusnw{\pgfkeysvalueof{/pgf/rectangle corner radius north west}}%
  }
  \savedmacro\cornerradiusne{%
    \edef\cornerradiusne{\pgfkeysvalueof{/pgf/rectangle corner radius north east}}%
  }
  \savedmacro\cornerradiussw{%
    \edef\cornerradiussw{\pgfkeysvalueof{/pgf/rectangle corner radius south west}}%
  }
  \savedmacro\cornerradiusse{%
    \edef\cornerradiusse{\pgfkeysvalueof{/pgf/rectangle corner radius south east}}%
  }

  \backgroundpath{%
    \northeast\advance\pgf@y-\cornerradiusne\relax
    \pgfpathmoveto{}%
    \pgfpatharc{0}{90}{\cornerradiusne}%
    \northeast\pgf@ya=\pgf@y\southwest\advance\pgf@x\cornerradiusnw\relax\pgf@y=\pgf@ya
    \pgfpathlineto{}%
    \pgfpatharc{90}{180}{\cornerradiusnw}%
    \southwest\advance\pgf@y\cornerradiussw\relax
    \pgfpathlineto{}%
    \pgfpatharc{180}{270}{\cornerradiussw}%
    \northeast\pgf@xa=\pgf@x\advance\pgf@xa-\cornerradiusse\southwest\pgf@x=\pgf@xa
    \pgfpathlineto{}%
    \pgfpatharc{270}{360}{\cornerradiusse}%
    \northeast\advance\pgf@y-\cornerradiusne\relax
    \pgfpathlineto{}%
    \pgfpathclose
  }

  \anchor{before north east}{\northeast\advance\pgf@y-\cornerradiusne}
  \anchor{after north east}{\northeast\advance\pgf@x-\cornerradiusne}
  \anchor{before north west}{\southwest\pgf@xa=\pgf@x\advance\pgf@xa\cornerradiusnw
    \northeast\pgf@x=\pgf@xa}
  \anchor{after north west}{\northeast\pgf@ya=\pgf@y\advance\pgf@ya-\cornerradiusnw
    \southwest\pgf@y=\pgf@ya}
  \anchor{before south west}{\southwest\advance\pgf@y\cornerradiussw}
  \anchor{after south west}{\southwest\advance\pgf@x\cornerradiussw}
  \anchor{before south east}{\northeast\pgf@xa=\pgf@x\advance\pgf@xa-\cornerradiusse
    \southwest\pgf@x=\pgf@xa}
  \anchor{after south east}{\southwest\pgf@ya=\pgf@y\advance\pgf@ya\cornerradiusse
    \northeast\pgf@y=\pgf@ya}

  \anchorborder{%
    \pgf@xb=\pgf@x%
    \pgf@yb=\pgf@y%
    \southwest%
    \pgf@xa=\pgf@x%
    \pgf@ya=\pgf@y%
    \northeast%
    \advance\pgf@x by-\pgf@xa%
    \advance\pgf@y by-\pgf@ya%
    \pgf@xc=.5\pgf@x%
    \pgf@yc=.5\pgf@y%
    \advance\pgf@xa by\pgf@xc%
    \advance\pgf@ya by\pgf@yc%
    \edef\pgf@marshal{%
      \noexpand\pgfpointborderrectangle
      {\noexpand\pgfqpoint{\the\pgf@xb}{\the\pgf@yb}}
      {\noexpand\pgfqpoint{\the\pgf@xc}{\the\pgf@yc}}%
    }%
    \pgf@process{\pgf@marshal}%
    \advance\pgf@x by\pgf@xa%
    \advance\pgf@y by\pgf@ya%
    \pgfextract@process\borderpoint{}%
    \pgf@rectanglewrc@donecorner@false
    \southwest\pgf@xc=\pgf@x\pgf@yc=\pgf@y
    \advance\pgf@xc\cornerradiussw\relax\advance\pgf@yc\cornerradiussw\relax
    \borderpoint
    \ifdim\pgf@x<\pgf@xc\relax\ifdim\pgf@y<\pgf@yc\relax
      \pgf@rectanglewithroundedcorners@docorner{-\cornerradiussw}{0pt}{180}{270}{\cornerradiussw}%
    \fi\fi
    \ifpgf@rectanglewrc@donecorner@\else
      \southwest\pgf@yc=\pgf@y\relax\northeast\pgf@xc=\pgf@x\relax
      \advance\pgf@xc-\cornerradiusse\relax\advance\pgf@yc\cornerradiusse\relax
      \borderpoint
      \ifdim\pgf@x>\pgf@xc\relax\ifdim\pgf@y<\pgf@yc\relax
       \pgf@rectanglewithroundedcorners@docorner{0pt}{-\cornerradiusse}{270}{360}{\cornerradiusse}%
      \fi\fi
    \fi
    \ifpgf@rectanglewrc@donecorner@\else
      \northeast\pgf@xc=\pgf@x\relax\pgf@yc=\pgf@y\relax
      \advance\pgf@xc-\cornerradiusne\relax\advance\pgf@yc-\cornerradiusne\relax
      \borderpoint
      \ifdim\pgf@x>\pgf@xc\relax\ifdim\pgf@y>\pgf@yc\relax
       \pgf@rectanglewithroundedcorners@docorner{\cornerradiusne}{0pt}{0}{90}{\cornerradiusne}%
      \fi\fi
    \fi
    \ifpgf@rectanglewrc@donecorner@\else
      \northeast\pgf@yc=\pgf@y\relax\southwest\pgf@xc=\pgf@x\relax
      \advance\pgf@xc\cornerradiusnw\relax\advance\pgf@yc-\cornerradiusnw\relax
      \borderpoint
      \ifdim\pgf@x<\pgf@xc\relax\ifdim\pgf@y>\pgf@yc\relax
       \pgf@rectanglewithroundedcorners@docorner{0pt}{\cornerradiusnw}{90}{180}{\cornerradiusnw}%
      \fi\fi
    \fi
  }
}

\makeatother

\tikzset{prepare tall/.style={rectangle with rounded corners, rectangle corner radius north east=0pt, rectangle corner radius south east=0pt, rectangle corner radius north west=30pt, rectangle corner radius south west=30pt, color=black,draw=black,fill=white,thick,inner sep=1pt,minimum height=22mm,minimum width=12mm}}
\tikzset{measure tall/.style={rectangle with rounded corners, rectangle corner radius north west=0pt, rectangle corner radius south west=0pt, rectangle corner radius north east=30pt, rectangle corner radius south east=30pt, color=black,draw=black,fill=white,thick,inner sep=1pt,minimum height=22mm,minimum width=12mm}}

%% file: tikzicons.tex
\definecolor{c666666}{RGB}{102,102,102}

\tikzset{
  server/.pic={
  \begin{scope}[scale=0.1]
      
  \path[fill=c666666,even odd rule] (27.8, -30.1) -- (28.6, -30.1) -- (28.6, 
  -31.2) -- (27.8, -31.2) -- cycle;

  \path[fill=c666666,even odd rule] (27.8, -31.8) -- (28.6, -31.8) -- (28.6, 
  -32.6) -- (27.8, -32.6) -- cycle;

  \path[fill=c666666,even odd rule] (29.2, -31.8) -- (33.3, -31.8) -- (33.3, 
  -32.6) -- (29.2, -32.6) -- cycle;

  \path[fill=c666666,even odd rule] (23.1, -31.8) -- (27.2, -31.8) -- (27.2, 
  -32.6) -- (23.1, -32.6) -- cycle;

\foreach \y in {0,2.8,5.6}{
    \begin{scope}[shift={(0,\y)}]
        
      \path[fill=c666666,even odd rule] (32.2, -27.8) -- (24.2, -27.8).. controls 
      (23.9, -27.8) and (23.7, -28.0) .. (23.7, -28.3) -- (23.7, -29.6).. controls 
      (23.7, -29.8) and (23.9, -30.1) .. (24.2, -30.1) -- (32.2, -30.1).. controls 
      (32.5, -30.1) and (32.7, -29.8) .. (32.7, -29.6) -- (32.7, -28.3).. controls 
      (32.7, -28.0) and (32.5, -27.8) .. (32.2, -27.8) -- cycle(25.1, -29.2).. 
      controls (24.9, -29.2) and (24.8, -29.1) .. (24.8, -28.9).. controls (24.8, 
      -28.8) and (24.9, -28.6) .. (25.1, -28.6).. controls (25.3, -28.6) and (25.4, 
      -28.8) .. (25.4, -28.9).. controls (25.4, -29.1) and (25.3, -29.2) .. (25.1, 
      -29.2) -- cycle(26.5, -29.2).. controls (26.4, -29.2) and (26.2, -29.1) .. 
      (26.2, -28.9).. controls (26.2, -28.8) and (26.4, -28.6) .. (26.5, -28.6).. 
      controls (26.7, -28.6) and (26.8, -28.8) .. (26.8, -28.9).. controls (26.8, 
      -29.1) and (26.7, -29.2) .. (26.5, -29.2) -- cycle(27.9, -29.2).. controls 
      (27.8, -29.2) and (27.6, -29.1) .. (27.6, -28.9).. controls (27.6, -28.8) and 
      (27.8, -28.6) .. (27.9, -28.6).. controls (28.1, -28.6) and (28.2, -28.8) .. 
      (28.2, -28.9).. controls (28.2, -29.1) and (28.1, -29.2) .. (27.9, -29.2) -- 
      cycle;
    \end{scope}
  }
  \end{scope}
  }
}

\definecolor{c6b9ac4}{RGB}{107,154,196}

\tikzset{
  file/.pic={
  \begin{scope}[scale=0.1]
  \path[fill=c6b9ac4,even odd rule] (92.2, -38.9) -- (92.2, -33.3) -- (94.3, 
  -33.3) -- (94.3, -35.0) -- (96.3, -35.0) -- (96.3, -38.9) -- cycle(94.7, 
  -33.5) -- (95.8, -34.5) -- (94.7, -34.5) -- cycle(94.7, -32.8) -- (91.7, 
  -32.8) -- (91.7, -39.4) -- (96.8, -39.4) -- (96.8, -34.6) -- cycle;

  \path[fill=c6b9ac4,even odd rule] (92.7, -35.8) -- (95.8, -35.8) -- (95.8, 
  -36.2) -- (92.7, -36.2) -- cycle;

  \path[fill=c6b9ac4,even odd rule] (92.7, -35.2) -- (93.7, -35.2) -- (93.7, 
  -35.5) -- (92.7, -35.5) -- cycle;

  \path[fill=c6b9ac4,even odd rule] (92.7, -36.5) -- (95.8, -36.5) -- (95.8, 
  -36.8) -- (92.7, -36.8) -- cycle;

  \path[fill=c6b9ac4,even odd rule] (92.7, -37.2) -- (95.8, -37.2) -- (95.8, 
  -37.5) -- (92.7, -37.5) -- cycle;

  \path[fill=c6b9ac4,even odd rule] (92.7, -37.8) -- (95.8, -37.8) -- (95.8, 
  -38.2) -- (92.7, -38.2) -- cycle;
  \end{scope}
}}